\documentclass[prb,aps,twocolumn]{revtex4}
\usepackage{epsfig,epsf,psfrag}
\usepackage{graphicx}
\usepackage{epic,eepic}
\usepackage{color,pstcol}
\begin{document}
\title{Gate-tunable zero-frequency current cross-correlations of the
  quartet mode in a voltage-biased three-terminal Josephson junction}

\date{\today}
\author{R\'egis M\'elin}
\affiliation{Universit\'e Grenoble-Alpes, Institut N\'eel, BP
  166, F-38042 Grenoble Cedex 9, France}
\affiliation{CNRS, Institut N\'eel, BP 166, F-38042 Grenoble
  Cedex 9, France}

\author{Mo\"{\i}se Sotto}
\affiliation{Universit\'e Grenoble-Alpes, Institut N\'eel, BP
  166, F-38042 Grenoble Cedex 9, France}
\affiliation{CNRS, Institut N\'eel, BP 166, F-38042 Grenoble
  Cedex 9, France}

\author{Denis Feinberg}
\affiliation{Universit\'e Grenoble-Alpes, Institut N\'eel, BP
  166, F-38042 Grenoble Cedex 9, France}
\affiliation{CNRS, Institut N\'eel, BP 166, F-38042 Grenoble
  Cedex 9, France}

\author{Jean-Guy Caputo}
\affiliation{Laboratoire de Math\'ematiques, INSA de Rouen,
  Avenue de l'Universit\'e, F-76801 Saint-Etienne du Rouvray, France}

\author{Beno\^{\i}t Dou\c{c}ot} \affiliation{Laboratoire de Physique
  Th\'eorique et des Hautes Energies, CNRS UMR 7589, Universit\'e
  Pierre et Marie Curie, Sorbonne Universit\'es, 4 Place Jussieu,
  75252 Paris Cedex 05}

\begin{abstract}
A three-terminal Josephson junction biased at opposite voltages can
sustain a phase-sensitive dc-current carrying three-body static phase
coherence, known as the ``quartet current''. We calculate the
zero-frequency current noise cross-correlations and answer the
question of whether this current is noisy (like a normal current in
response to a voltage drop) or noiseless (like an equilibrium
supercurrent in response to a phase drop). A quantum dot with a level
at energy $\epsilon_0$ is connected to three superconductors $S_a$,
$S_b$ and $S_c$ with gap $\Delta$, biased at $V_a=V$, $V_b=-V$ and
$V_c=0$, and with intermediate contact transparencies. At zero
temperature, nonlocal quartets (in the sense of four-fermion
correlations) are noiseless at subgap voltage in the nonresonant dot
regime $\epsilon_0/\Delta\gg 1$, which is demonstrated with a
semi-analytical perturbative expansion of the cross-correlations.
Noise reveals the absence of granularity of the superflow
  splitting from $S_c$ towards $(S_a,S_b)$ in the nonresonant dot regime,
in spite of finite voltage. In the resonant dot regime
$\epsilon_0/\Delta\alt 1$, cross-correlations measured in the
$(V_a,V_b)$ plane should reveal an ``anomaly'' in the vicinity of the
quartet line $V_a+V_b=0$, related to an additional contribution to the
noise, manifesting the phase sensitivity of cross-correlations
under the appearance of a three-body phase
variable. Phase-dependent effective Fano factors
$F_\varphi$ are introduced, defined as the ratio between the
amplitudes of phase modulations of the noise and the currents. At low
bias, the Fano factors $F_\varphi$ are of order unity in the resonant dot
regime $\epsilon_0/\Delta\alt 1$, and they are vanishingly small in
the nonresonant dot regime $\epsilon_0/\Delta\gg 1$.
\end{abstract}
\maketitle

\section{Introduction}
The Josephson effect and multiple Andreev reflections (MARs) appear to
be well established at present time in two-terminal set-ups
\cite{Averin1,Averin2,Cuevas,Cuevas-noise}, especially with respect to
the clearcut break-junction experiments\cite{Saclay1,Saclay2}. Less is
known about three terminals. A few recent works
\cite{Cuevas-Pothier,Freyn,Jonckheere,Francois,Houzet-Samuelsson,Duhot,Riwar,Padurariu,Delft}
dealt with superconducting nanoscale devices with three
superconductors $S_a$, $S_b$ and $S_c$ biased at $V_a$, $V_b$ and
$V_c=0$, instead of superconducting weak links with only two
terminals. It was established by Cuevas and Pothier
\cite{Cuevas-Pothier} on the basis of Usadel equations that the third
terminal $S_c$ can be viewed qualitatively as having the same effect
as an rf-source, producing what was coined \cite{Cuevas-Pothier} as
``self-induced Shapiro steps''. Later, Freyn {\it et al.}\cite{Freyn}
rediscovered those {voltage} resonances, and identified in the adiabatic regime
the emergence of intermediate states involving correlations among
four, six, eight,~... fermions (the so-called quartets, sextets,
octets,~...). The condition for appearance of a coherent dc-current at
a $(p,q)$-resonance is $p(V_a-V_c)+q(V_b-V_c)=0$. Nonlocal quartets
correspond to $(p,q)=(1,1)$, nonlocal sextets to $(p,q)=(1,2)$ or
$(2,1)$, nonlocal octets to $(p,q)=(1,3)$, $(2,2)$ or $(3,1)$,~... For
tunnel contacts and at low bias, allowing an adiabatic approximation,
the dc-current at a $(p,q)$ resonance is given by{
\begin{eqnarray}
\label{eq1}
\nonumber
&&I^c_{p,q}
\sin\left[p\left(\varphi_a(t)-\varphi_c(t)\right)
  +q\left(\varphi_b(t)-\varphi_c(t)\right)\right]\\
 &=&
I^c_{p,q} \sin\left[p\left(\varphi_a(0)-\varphi_c(0)\right)+
  q\left(\varphi_b(0)-\varphi_c(0)\right)\right] ,
\label{eq3}
\end{eqnarray}
where the last identity is valid only at a $(p,q)$ resonance
$p(V_a-V_c)+q(V_b-V_c)=0$} at which the nonlocal Josephson effect
becomes time-independent, and the critical current $I^c_{p,q}$ can be
calculated at equilibrium. It turns out that the microscopic process
of four-fermion exchange produces a $\pi$-shifted current-phase
relation for the quartets \cite{Jonckheere} {instead of a standard
  ``$0$''-junction}. A fully nonequilibrium calculation for the
current at {voltage} resonance was carried out by Jonckheere {\it et
  al.}\cite{Jonckheere} Correlations among pairs and quasiparticles
were also obtained in this work in the form of phase-sensitive MARs
(ph-MARs).  The recent Grenoble experiment on metallic junctions by
Pfeffer {\it et al.}\cite{Francois} provided evidence for phenomena
compatible with quartets. {However, this experiment could not firmly
  establish whether the anomaly is due to the quartet mechanism or the
  oscillations of populations. This question is somehow marginal: both
  effects can appear simultaneously because of cross-over between
  those two limiting cases. An more relevant question is that of
  reconsidering synchronization in a phase-coherent mesoscopic
  sample.}

It is shown here that noise experiments in three-terminal Josephson
junctions {should} provide complementary characterization of
those phase-coherent processes, similarly to Cooper pair splitting in
three-terminal normal metal-superconductor-normal metal set-ups
\cite{ref5,ref6-7,Samuelsson,ref9,Faoro,Bignon,Morten,Melin-Martin,Freyn-Floser-Melin,Floser-Feinberg-Melin,Heiblum,Takis}. A
well-understood mechanism for noise in voltage-biased normal-fermionic
junctions is partition noise. A well-known intuitive picture envisions
a (noiseless) incoming beam of regularly spaced fermionic
wave-packets. Each wave-packet incoming on the barrier is transmitted
with probability $T$, and reflected with probability $1-T$. {The
  randomness of the transmission process} produces noise in the
transmitted signal. The noise at arbitrary transmission $T$ is
proportional to $T(1-T)$, to the charge of the carriers, and to the
voltage.

However, this physical picture for partition noise does not apply to
equilibrium superflows of Cooper pairs in response to different phases
on different leads (in the absence of applied voltage), because a
superflow is collective and nongranular. The calculations presented in
the main body of our article are based on a previous article by Cuevas
{\it et al.}\cite{Cuevas-noise}, which turned out to be successful in
establishing the noise of MARs in a two-terminal set-up. In
particular, a dc-Josephson current is noiseless at zero temperature.

Let us come back to three-terminals superconducting junctions, which
contain both ingredients of applied voltages and dc-supercurrent of
pairs \cite{Freyn}. Three-body static phase coherence is present in
both the pair current (quartets or multipairs) and the quasiparticle
current (multiple Andreev reflections, MARs). One expects that noise
cross-correlations may help to separate the underlying microscopic
processes. A natural question arises in view of the discussion above
on the noise of a normal or superconducting flow: Is the
phase-sensitive current noiseless or noisy? This question is the
subject of the {present} article.

The notion of ``quartets'' is now discussed from a different
perspective.  Focusing on the case {$V_a-V_c=-(V_b-V_c)$}, the
appearance of a Josephson-like dc-current between, on the one hand,
$S_c$, and on the other hand the pair ($S_a,S_b$), signals the
existence of {\it static} phase coherence between the three
superconductors, despite the presence of nonzero voltages, in the
absence of static phase coherence between two conductors only. The
three-body coherence manifests itself in the relevant
{''quartet''} three-body phase variable
$\varphi_Q=\varphi_a+\varphi_b-2\varphi_c$, that is in principle
controllable with superconducting loops. In general, the dc-current is
a periodic function of the variable $\varphi_Q$. The latter form
suggests that instead of exchanging single Cooper pairs as in a $SNS$
or $SIS$ junction ($I$ is an insulating barrier), the exchange
``currency'' to establish static phase coherence between $S_c$ and
($S_a,S_b$) is an electronic quartet. This macroscopic point of view
is valid irrespective of the nature of the {junction} and of the
parameter regime, close or far from equilibrium.

Two distinct pictures for the notion of ``quartets'' are then
envisioned. Notion A corresponds to the restrictive sense of {\it
  four-fermion correlations}, as those appearing in the adiabatic
regime \cite{Freyn}. The (more general) notion B is that of {\it a
  currency exchanged to establish three-body static phase coherence},
characterized by the {``quartet phase''} $\varphi_Q$. It will be
shown that the resonant dot regime $\Gamma/\Delta\sim 1$ and
$\epsilon_0/\Delta\alt 1$ leads to finite phase-sensitive noise for
the quartets according to B. [The parameter $\epsilon_0$ is the
  quantum dot energy level with respect to the chemical potential of
  lead $S_c$.] By contrast, nonresonant-dot quartets (for
$\epsilon_0/\Delta\gg 1$) corresponding to A will be shown to be
noiseless once the current cross-correlations will be normalized to
the currents. A gate voltage can be used to cross-over from the
nonresonant ($\epsilon_0/\Delta\gg 1$) to the resonant dot
($\epsilon_0/\Delta\alt 1$) regimes, and thus to control the value of
the noise in the quartet {mode}.

Further technical introductory material is presented in
Sec.~\ref{sec:intro-technique}. It will be shown in
Sec.~\ref{sec:perturbative-quartets} on the basis of semi-analytical
calculations that the nonresonant-dot quartet current looks like an
equilibrium dc-Josephson current in the sense that it is
noiseless. Sec.~\ref{sec:3T-Josephson-junction} demonstrates by
numerical calculations that finite noise and noise cross-correlations
are produced in the resonant dot regime, depending on the three-body phase
variable $\varphi_Q$ mentioned above. It will be concluded in
Sec.~\ref{sec:conclusions} that an anomaly in the noise or in the
noise cross-correlations may be observed in future experiments with
resonant quantum dots. Moreover, the anomaly in the noise is predicted
to disappear as the quantum dot energy level is made nonresonant by
applying a gate voltage, because of a cross-over towards a collective
nongranular flow of Cooper pairs in the presence of finite voltages.

\begin{figure}[htb]
\includegraphics[width=\columnwidth]{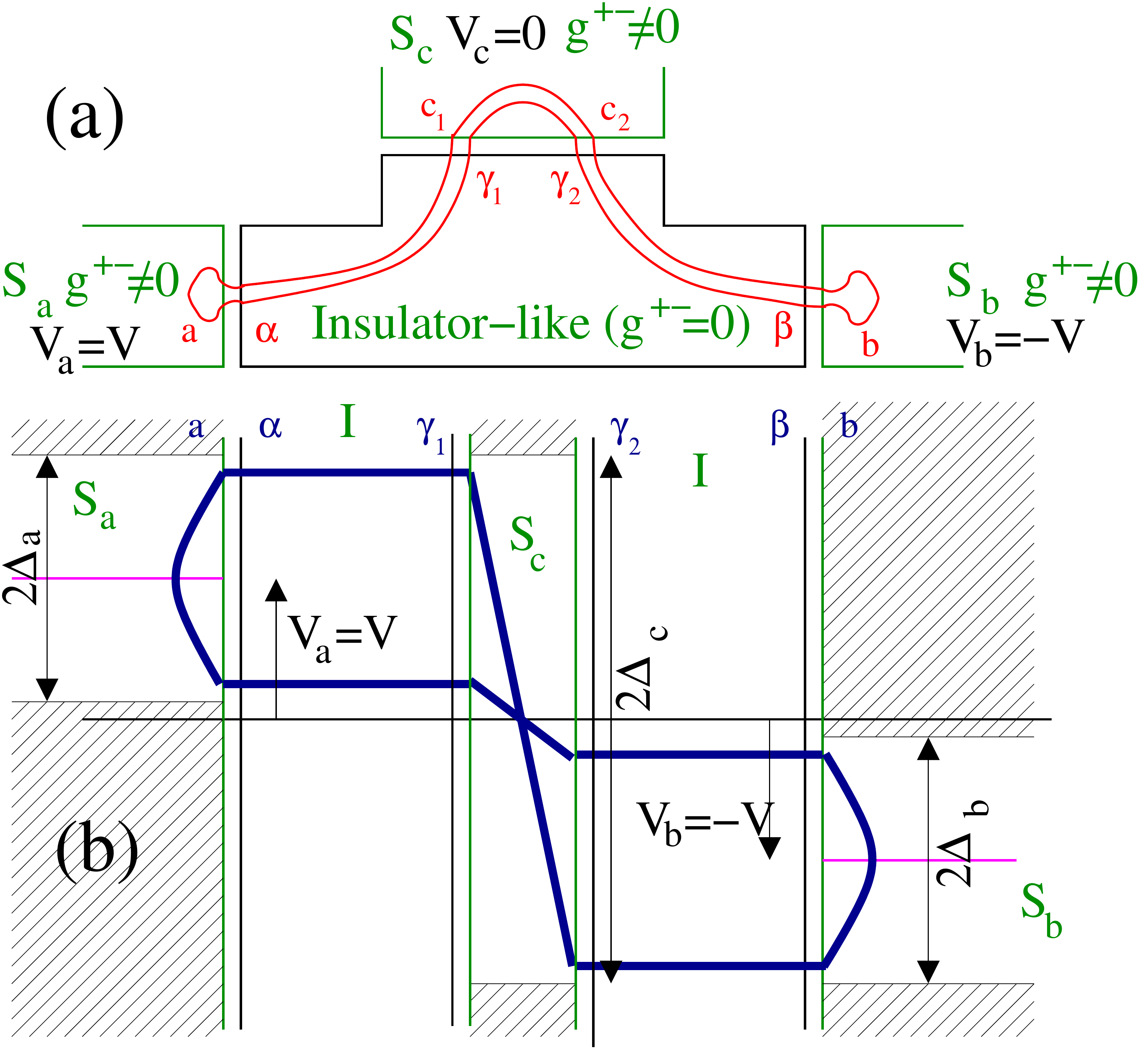}
\caption{The figure shows a set-up in which three superconductors
  $S_a$, $S_b$ and $S_c$ biased at $V_a=V$, $V_b=-V$ and $V_c=0$ are
  connected to a common insulator-like region. Panels a and b
  correspond respectively to space and energy representations for the
  butterfly diagram, encoding nonresonant-dot quartets with a current
  $I_c$ though $S_c$ set by $I_c=I_c^{(0)} \sin\varphi_Q$, with
  $\varphi_Q=\varphi_a+\varphi_b-2\varphi_c$. On this figure, the
  central region is equivalent to an insulator-like region, used to
  address the nonresonant dot regime in
  Sec.~\ref{sec:perturbative-quartets}. However, the numerical
  calculations in the forthcoming Sec.~\ref{sec:3T-Josephson-junction}
  deal with a set-up containing an embedded quantum dot (see
  Fig.~\ref{fig:quantum-dot}).
\label{fig:schema-isolant}
}
\end{figure}

\section{Expression of the noise in terms of Keldysh Green's functions}
\label{sec:intro-technique}
\subsection{Expression of the noise}
Three superconducting leads $S_a$, $S_b$ and $S_c$ biased at $V_a=V$,
$V_b=-V$ and $V_c=0$ are connected to a common region (insulating or
nonresonant quantum dot). The method used here is taken from the
papers by Cuevas {\it et al.}\cite{Cuevas,Cuevas-noise} on the current
and noise of a two-terminal superconducting contact (see the
Appendix).

The kernel of current-current correlations between terminals $a_k$ and
$a_l$ ($a_k,a_l\in\{S_a,S_b,S_c\}$) is given by
\begin{equation}
\label{eq:K}
K_{a_k,a_l}(\tau,\tau')=\langle \delta I_{a_k}(\tau+\tau') \delta
I_{a_l}(\tau)\rangle,
\end{equation}
where $\delta I_{a_k}(\tau)$ is the current fluctuation of terminal
$a_k$ at time $\tau$. Because of the explicit time-dependence of the
Hamiltonian, the current-current correlations
$S_{a_k,a_l}(\tau,\tau')$ depend on both times $\tau$ and $\tau'$, not
only on $\tau-\tau'$. The noise correlations are given
by\cite{Cuevas-noise}
\begin{equation}
S_{a_k,a_l}(\tau)=\hbar \int d\tau' K_{a_k,a_l}(\tau,\tau')
\end{equation}
The kernel given by Eq.~(\ref{eq:K}) is expressed in terms of the
Keldysh Green's functions $\hat{G}^{+,-}$ and $\hat{G}^{-,+}$, for
instance:
\begin{eqnarray}
\nonumber
&&\hat{K}_{a,b}(\tau,\tau')=\frac{e^2}{\hbar^2}\mbox{Tr}\\
\label{eq:Sterme1}
&&\left\{\hat{\Sigma}_{\beta,b}(\tau) \hat{\tau}_3
\hat{G}^{+,-}_{b,a}(\tau,\tau')
\hat{\Sigma}_{a,\alpha}(\tau')\hat{\tau}_3
\hat{G}^{-,+}_{\alpha,\beta}(\tau',\tau)\right.\\
\label{eq:Sterme2}
&+&
\hat{\Sigma}_{b,\beta}(\tau) \hat{\tau}_3
\hat{G}^{+,-}_{\beta,\alpha}(\tau,\tau')
\hat{\Sigma}_{\alpha,a}(\tau') \hat{\tau}_3
\hat{G}^{-,+}_{a,b}(\tau',\tau)\\
\label{eq:Sterme3}
&-& \hat{\Sigma}_{\beta,b}(\tau)
\hat{\tau}_3 \hat{G}^{+,-}_{b,\alpha}(\tau,\tau')
\hat{\Sigma}_{\alpha,a}(\tau') \hat{\tau}_3
\hat{G}^{-,+}_{a,\beta}(\tau',\tau)\\
\label{eq:Sterme4}
&-& \hat{\Sigma}_{b,\beta}(\tau)\hat{\tau}_3
\hat{G}^{+,-}_{\beta,a}(\tau,\tau') \hat{\Sigma}_{a,\alpha}(\tau')
\hat{\tau}_3 \hat{G}^{-,+}_{\alpha,b}(\tau',\tau)\\ &&\left.+ (\tau
\leftrightarrow \tau')\right\},
\end{eqnarray}
where the trace ``Tr'' is a summation over the Nambu labels. Latin
labels $a$, $b$, $c$ are used for the tight-binding sites in the
superconducting leads, and Greek labels $\alpha$, $\beta$ and
$\gamma$ are used for the insulating region. The label $x$ will be
used in Sec.~\ref{sec:3T-Josephson-junction} for a zero-dimensional
quantum dot (see also the Appendix). Notations like
$\hat{\Sigma}_{a,\alpha}$ or $\hat{\Sigma}_{\alpha,a}$ have the
meaning of the hopping amplitude for crossing the interface $S_aI$
in the direction $a\rightarrow \alpha$ or $\alpha\rightarrow a$
respectively.  The Keldysh Green's functions in
Eqs.~(\ref{eq:Sterme1})-(\ref{eq:Sterme4}) are given by
\begin{eqnarray}
\nonumber
&&\hat{G}^{+,-}_{i,j}(\tau,\tau')=\\
&&i\left(\begin{array}{cc}
\langle c_{j \uparrow}^+(\tau') c_{i \uparrow}(\tau)\rangle &
\langle c_{j \downarrow}(\tau')c_{i \uparrow}(\tau)\rangle\\
\langle c_{j \uparrow}^+(\tau') c_{i \downarrow}^+(\tau)\rangle &
\langle c_{j \downarrow}(\tau')c_{i \downarrow}^+(\tau)\rangle
\end{array}\right)
\end{eqnarray}
and
\begin{eqnarray}
\nonumber
&&\hat{G}^{-,+}_{i,j}(\tau,\tau')=\\
&&-i\left(\begin{array}{cc}
\langle c_{i \uparrow}(\tau) c_{j \uparrow}^+(\tau')\rangle &
\langle c_{i \uparrow}(\tau)c_{j \downarrow}(\tau')\rangle\\
\langle c_{i \downarrow}^+(\tau) c_{j \uparrow}^+(\tau')\rangle &
\langle c_{j \downarrow}^+(\tau)c_{j \downarrow}(\tau')\rangle
\end{array}\right)
.
\end{eqnarray}
The gauge is such that the tunnel terms (purely diagonal in Nambu) are
time-dependent:
\begin{eqnarray}
\Sigma_{a_k,\alpha_k}^{1,1} (t) &=&
\Sigma_{a_k,\alpha_k}^{1,1} \exp(i V_{a_k} t / \hbar)\\
\Sigma_{a_k,\alpha_k}^{2,2} (t) &=& \Sigma_{a_k,\alpha_k}^{2,2} \exp(-i
V_{a_k} t / \hbar),
\end{eqnarray}
with $\Sigma_{\alpha_k,a_k}^{1,1} (t)=[\Sigma_{a_k,\alpha_k}^{1,1}
  (t)]^*$, and $\Sigma_{\alpha_k,a_k}^{2,2}
(t)=[\Sigma_{a_k,\alpha_k}^{2,2} (t)]^*$. The expression for the noise
kernel is conveniently Fourier transformed.

\subsection{Adiabatic limit}
\label{sec:adiabatic}
Of particular interest is to show that the noise vanishes in the
adiabatic limit. The adiabatic limit corresponds to very small applied
voltage, whatever interface transparencies. Then, the phases evolve
slowly in time and the Keldysh Green's functions are approximated as
being parameterized by the quasi-static phase variables
$\varphi_{a,b,c}$. Fourier transforming from the time difference
$\tau-\tau'$ to frequency $\omega$ leads to the following expression
for the Keldysh Green's functions:
\begin{eqnarray}
\label{eq:r1}
\hat{G}^{+,-}(\omega)&=&n_F(\omega)\left[G^A(\omega)-G^R(\omega)\right]\\
\hat{G}^{-,+}(\omega)&=&\left(n_F(\omega)-1\right)
\left[G^A(\omega)-G^R(\omega)\right]
,
\label{eq:r2}
\end{eqnarray}
where $n_F(\omega)$ is the equilibrium Fermi distribution function at
zero temperature, and at energy $\omega$. Those expressions are easily
deduced from the corresponding Dyson equations for the Keldysh Green's
function, which, in a compact notation, take the following form:
\begin{eqnarray}
\hat{G}^{+,-}&=&\left(\hat{I}+\hat{G}^R \hat{\Sigma}\right)\hat{g}^{+,-}
\left(\hat{I}+\hat{\Sigma}\hat{G}^A\right)\\
\label{eq:z1}
&=&\left(\hat{I}+\hat{G}^R \hat{\Sigma}\right)
n_F\left(\hat{g}^A-\hat{g}^R\right)
\left(\hat{I}+\hat{\Sigma}\hat{G}^A\right)\\
\label{eq:z2}
&=&n_F\left(\hat{I}+\hat{G}^R \hat{\Sigma}\right)
\left(\hat{g}^A-\hat{g}^R\right)
\left(\hat{I}+\hat{\Sigma}\hat{G}^A\right)\\
&=&n_F\left\{\left(\hat{I}+\hat{G}^R\hat{\Sigma}\right) \hat{G}^A
-\hat{G}^R\left(\hat{I}+\hat{\Sigma}\hat{G}^A\right)\right\}\\
&=&n_F\left(G^A(\omega)-G^R(\omega)\right)
\label{eq:z3}
.
\end{eqnarray}
Going from Eq.~(\ref{eq:z1}) to Eq.~(\ref{eq:z2}), it was used that
the voltages are identical in all leads, from what it is deduced that
the occupation number $n_F\equiv n_F(\omega)$ can be factored out.  It is
noticed that the noise kernel [see
  Eqs.~(\ref{eq:Sterme1})-(\ref{eq:Sterme4})] involves products
between Eqs.~(\ref{eq:r1}) and~(\ref{eq:r2}). Again, the Fermi
occupation numbers factor out, and the product
$n_F(\omega)\left(1-n_F(\omega)\right)$ is vanishingly small at zero
temperature: the current-current (cross-)correlations are vanishingly
small in the adiabatic limit.

The next step, considered in the following
Sec.~\ref{sec:perturbative-quartets}, is to show that the quartet
contribution to the noise cross-correlations is vanishingly small in
the nonresonant dot limit, for subgap voltages. The nonresonant dot limit
corresponds to very low interface transparencies, whatever {the} applied
voltages.

\section{Strongly nonresonant dot regime}
\label{sec:perturbative-quartets}
It is supposed in this section on perturbative calculations in
transparency that three superconducting leads $S_a,S_b$ and $S_c$ are
connected to a (small) common insulating region (see
Fig.~\ref{fig:schema-isolant}). This set-up on
Fig.~\ref{fig:schema-isolant} is equivalent to a strongly nonresonant
quantum dot (with $\epsilon_0/\Delta\gg 1$) embedded in a structure
with three superconductors. The strongly nonresonant regime
$\epsilon_0/\Delta\gg 1$ is addressed here with perturbation theory in
the junction transparency. The bare Keldysh Green's functions denoted
by $\hat{g}_{a,a}^{+,-}$, $\hat{g}_{b,b}^{+,-}$ and
$\hat{g}_{c,c}^{+,-}$ are finite in the superconducting leads, but the
bare Keldysh Green's function is vanishingly small in the insulator,
due to the absence of density of states in this
region\cite{Caroli}. An expansion of the current and noise in powers
of the tunnel amplitudes can be represented schematically by
diagrams. Of particular interest here is the ``butterfly diagram'' for
the quartets \cite{Freyn}, which forms a closed loop in space and in
energy (thus leading to a dc-term in the current and noise). The
microscopic process of quartets is the lowest order coupling to the
three-body phase variable $\varphi_Q$ (see
Fig.~\ref{fig:schema-isolant}). The calculation proceeds by expanding
each term contributing to the noise cross-correlations $S_{a,b}$ [see
  Eqs.~(\ref{eq:Sterme1})-(\ref{eq:Sterme4})] to order $\Sigma^8$
according to the quartet butterfly diagram. In addition, the Nambu
labels for electrons and holes are selected in such a way as to
produce the correct electron-hole conversions with respect to the
quartet butterfly diagram [see Fig.~\ref{fig:schema-isolant}b]. For
instance the ``11'' Nambu component of the term (\ref{eq:Sterme1}) is
given at order $\Sigma^8$ by the following three terms:
\begin{widetext}
\begin{eqnarray}
\label{eq:pert1}
&&\hat{\Sigma}_{\beta,b}^{1,1/0,1} \hat{\tau}_3^{1,1/1,1}
\hat{g}^{+,-/1,2/1,1}_{b,b} \hat{\Sigma}_{b,\beta}^{2,2/1,2}
\hat{g}^{A/2,2/2,2}_{\beta,\gamma_2}
\hat{\Sigma}^{2,2/2,2}_{\gamma_2,c_2} \hat{g}^{A/2,1/2,2}_{c_2,c_1}
\hat{\Sigma}^{1,1/2,2}_{c_1,\gamma_1}
\hat{g}^{A/1,1/2,2}_{\gamma_1,\alpha}\\ \nonumber&&\times
\hat{\Sigma}_{\alpha,a}^{1,1/2,1} \hat{g}^{A,1,2/1,1}_{a,a}
\hat{\Sigma}^{2,2/1,0}_{a,\alpha} \hat{\tau}_3^{2,2/0,0}
\hat{g}^{R/2,2/0,0}_{\alpha,\gamma_1}
\hat{\Sigma}^{2,2/0,0}_{\gamma_1,c_1} \hat{g}^{-,+/2,1/0,0}_{c_1,c_2}
\hat{\Sigma}^{1,1/0,0}_{c_2,\gamma_2}
\hat{g}^{A/1,1/0,0}_{\gamma_2,\beta} \\ &+&
\hat{\Sigma}_{\beta,b}^{1,1/0,1} \hat{\tau}_3^{1,1/1,1}
\hat{g}^{R/1,2/1,1}_{b,b} \hat{\Sigma}_{b,\beta}^{2,2/1,2}
\hat{g}^{R/2,2/2,2}_{\beta,\gamma_2}
\hat{\Sigma}^{2,2/2,2}_{\gamma_2,c_2} \hat{g}^{+,-/2,1/2,2}_{c_2,c_1}
\hat{\Sigma}^{1,1/2,2}_{c_1,\gamma_1}
\hat{g}^{A/1,1/2,2}_{\gamma_1,\alpha}\\\nonumber
&&\times \hat{\Sigma}_{\alpha,a}^{1,1/2,1} \hat{g}^{A/1,2/1,1}_{a,a}
\hat{\Sigma}^{2,2/1,0}_{a,\alpha} \hat{\tau}_3^{2,2/0,0}
\hat{g}^{R/2,2/0,0}_{\alpha,\gamma_1}
\hat{\Sigma}^{2,2/0,0}_{\gamma_1,c_1} \hat{g}^{-,+/2,1/0,0}_{c_1,c_2}
\hat{\Sigma}^{1,1/0,0}_{c_2,\gamma_2}
\hat{g}^{A/1,1/0,0}_{\gamma_2,\beta} \\ &+&
\hat{\Sigma}_{\beta,b}^{1,1/0,1} \hat{\tau}_3^{1,1/1,1}
\hat{g}^{R/1,2/1,1}_{b,b} \hat{\Sigma}_{b,\beta}^{2,2/1,2}
\hat{g}^{R/2,2/2,2}_{\beta,\gamma_2}
\hat{\Sigma}^{2,2/2,2}_{\gamma_2,c_2} \hat{g}^{R/2,1/2,2}_{c_2,c_1}
\hat{\Sigma}^{1,1/2,2}_{c_1,\gamma_1}
\hat{g}^{R/1,1/2,2}_{\gamma_1,\alpha}\\\nonumber
&&\times \hat{\Sigma}_{\alpha,a}^{1,1/2,1} \hat{g}^{+,-,1,2/1,1}_{a,a}
\hat{\Sigma}^{2,2/1,0}_{a,\alpha} \hat{\tau}_3^{2,2/0,0}
\hat{g}^{R/2,2/0,0}_{\alpha,\gamma_1}
\hat{\Sigma}^{2,2/0,0}_{\gamma_1,c_1} \hat{g}^{-,+/2,1/0,0}_{c_1,c_2}
\hat{\Sigma}^{1,1/0,0}_{c_2,\gamma_2}
\hat{g}^{A/1,1/0,0}_{\gamma_2,\beta}
\label{eq:pert3}
,
\end{eqnarray}
\end{widetext}
and similar expressions are obtained for all of the $28$ terms
contributing to the current-current cross-correlations [see
  Eqs.~(\ref{eq:Sterme1})-(\ref{eq:Sterme4})].  Expressions like
$\hat{\Sigma}_{a,\alpha}^{\tau_1,\tau_2/n_1,n_2}$ have the meaning of
traversing the interface $S_aI$ upon changing the Nambu labels
according to $\tau_1\rightarrow \tau_2$, and the labels of harmonics
according to $n_1\rightarrow n_2$. The hopping amplitudes do not
change the value $\tau_2=\tau_1$ of the Nambu labels, but they
increment by $n_2=n_1 \pm 1$ the label of harmonics. On the contrary,
the anomalous bare Green's function changes the value of the Nambu
labels, but the labels of harmonics are left unchanged because of the
choice of the gauge.

\begin{figure}[htb]
\includegraphics[width=.8\columnwidth]{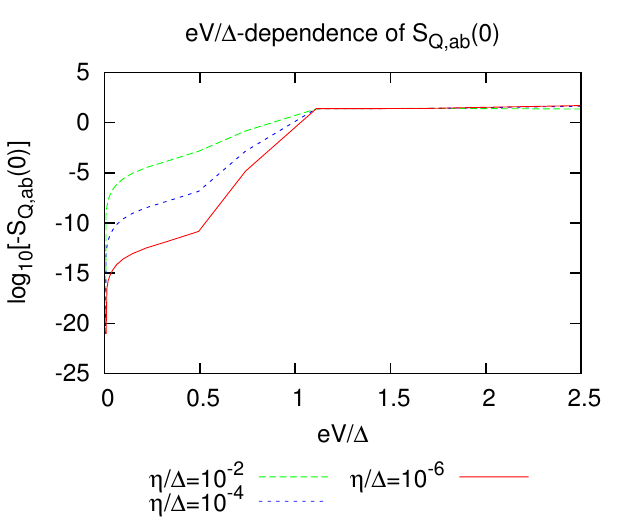}
\caption{Voltage dependence of the quartet contribution to the
  current-current cross-correlations, for the values of $\eta/\Delta$
  shown on the figure. The parameter $\eta$ is a regulator introduced
  as the imaginary part of the energy $\omega$.
\label{fig:pert-dep-V}
}
\end{figure}

\begin{figure}[htb]
\includegraphics[width=.6\columnwidth]{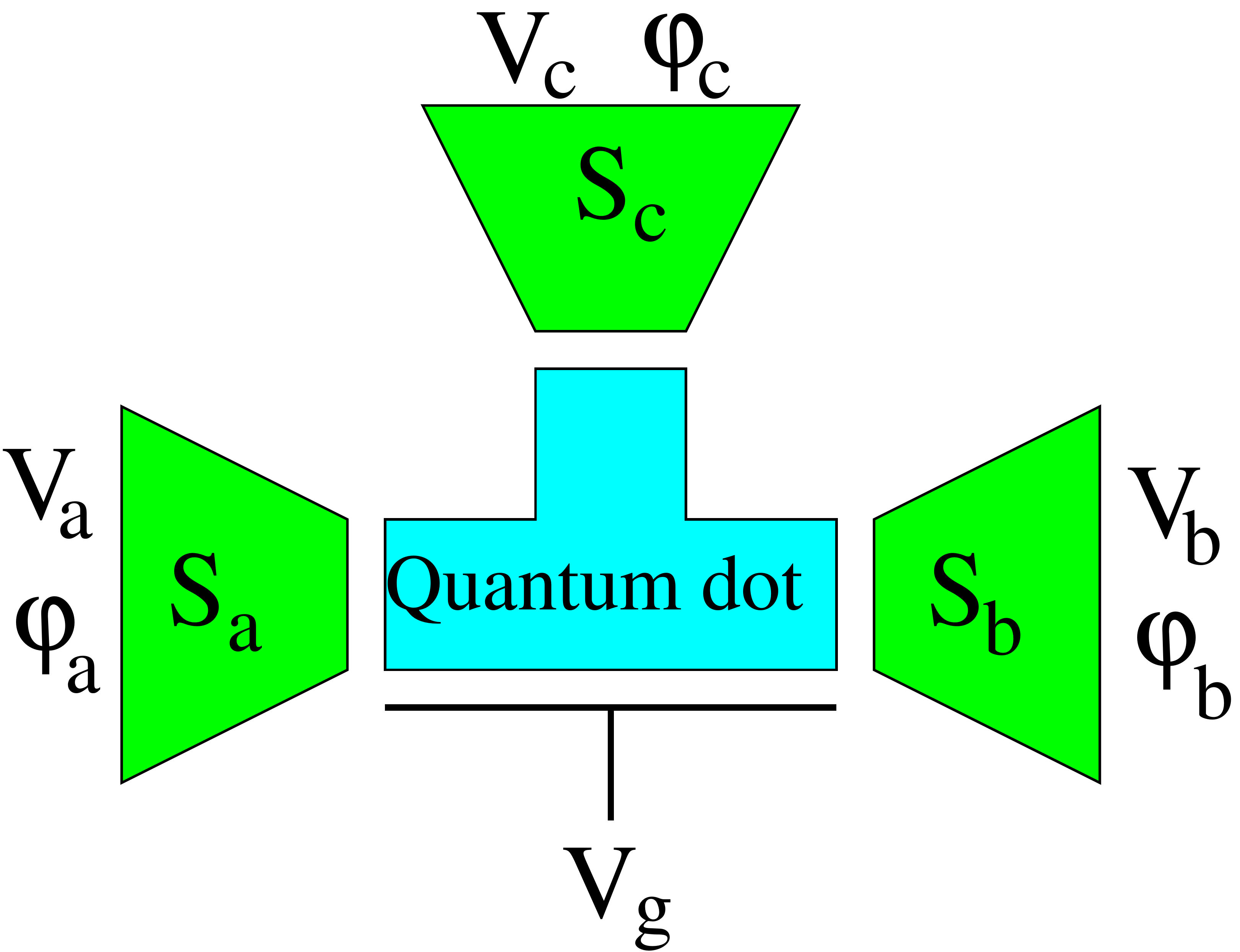}
\caption{Schematics of a quantum dot connected to three
  superconductors $S_a$, $S_b$ and $S_c$ at voltages $V_a$, $V_b$ and
  $V_c$. A gate voltage $V_g$ is applied to the quantum dot.
\label{fig:quantum-dot}}
\end{figure}

It is first shown that our expansion in the tunnel amplitude $\Sigma$
is compatible with the vanishingly small value of the noise in the
adiabatic limit (see Sec.~\ref{sec:intro-technique}). For this
purpose, we collected the only four terms at order $\Sigma^8$
containing only advanced or only retarded Green's functions, but no
products between the former and the latter. It is indeed those terms
that encode the adiabatic limit, because the current in this limit is
expressed as the sum or difference of terms that contain only advanced
or only retarded Green's functions [see the form of the Keldysh
  Green's function in Eq.~(\ref{eq:z3})]. Once those terms are
identified, it is easy to show for the harmonics labels that the
Green's functions $\hat{g}^{+,-}_{a,a}$ and $\hat{g}^{+,-}_{b,b}$
contain identical sets of harmonics labels, meaning that those terms
do not contribute to the noise at zero temperature, because of a
prefactor of the type $n_F(\omega + p \omega_0 / 2) [n_F(\omega + p
  \omega_0 / 2 ) - 1]$, where $p$ is an integer. The contribution of
those {``adiabatic''} terms to the noise is thus vanishingly
small, in agreement with the discussion of the adiabatic limit in
Sec.~\ref{sec:adiabatic}.

\begin{figure*}[htb]
{
\centerline{\includegraphics[width=.5\textwidth]{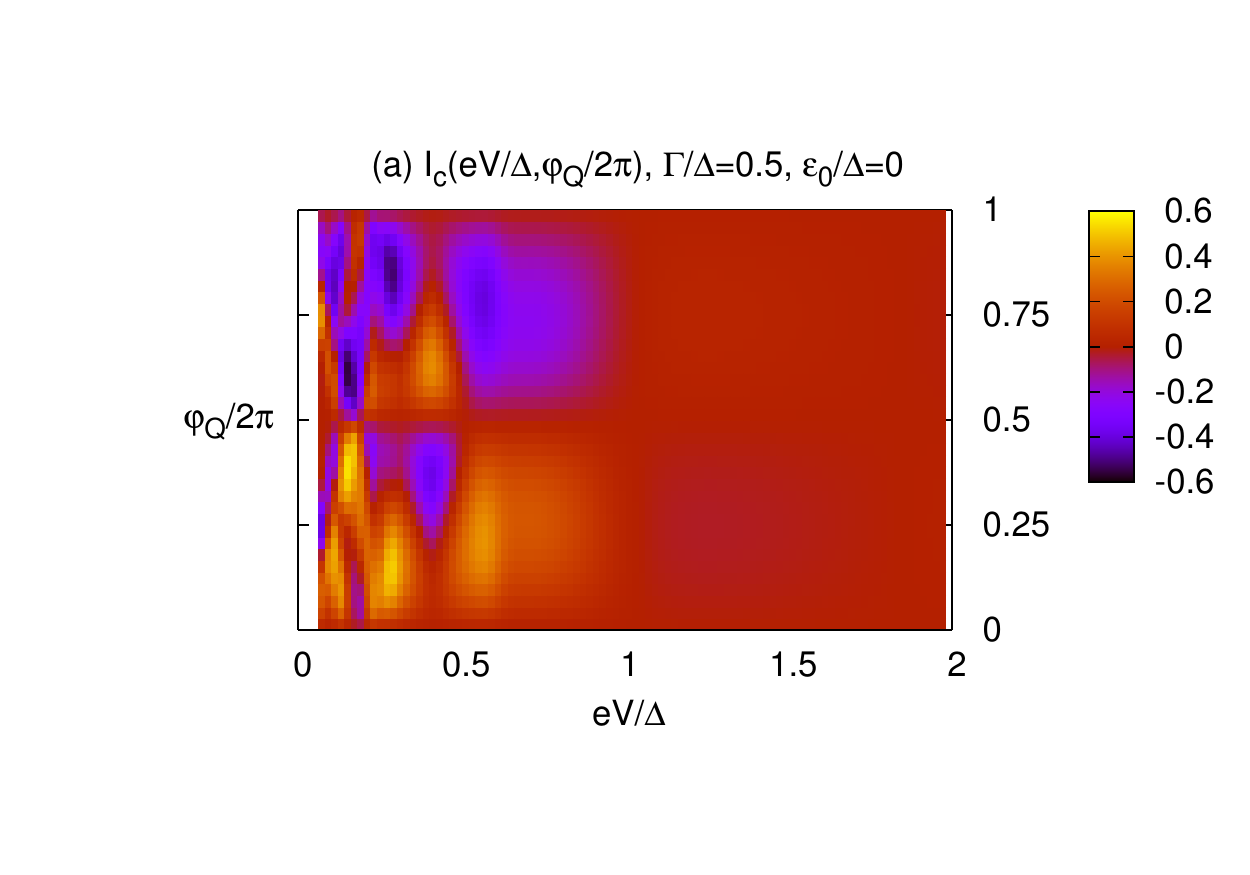}\includegraphics[width=.5\textwidth]{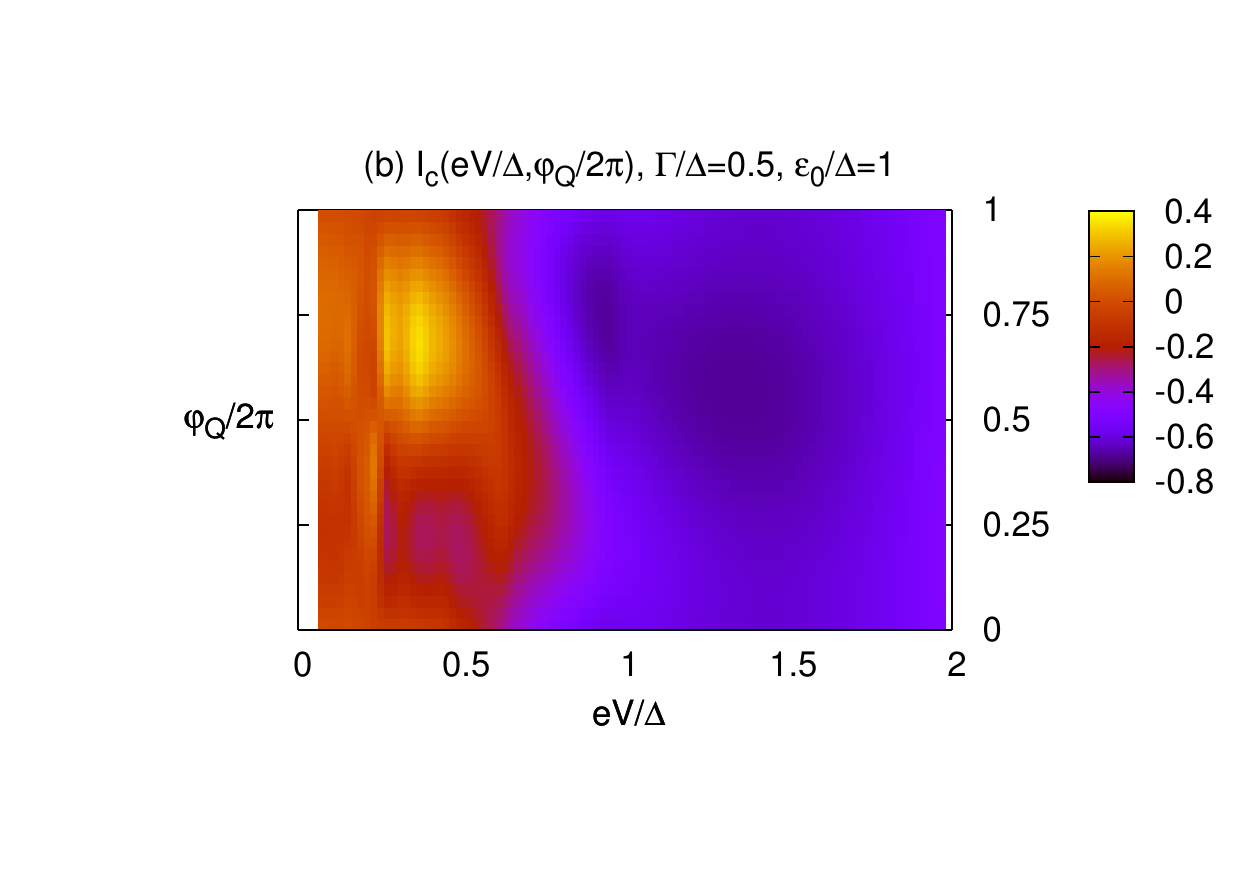}}
\vspace*{-1.5cm}
\centerline{\includegraphics[width=.5\textwidth]{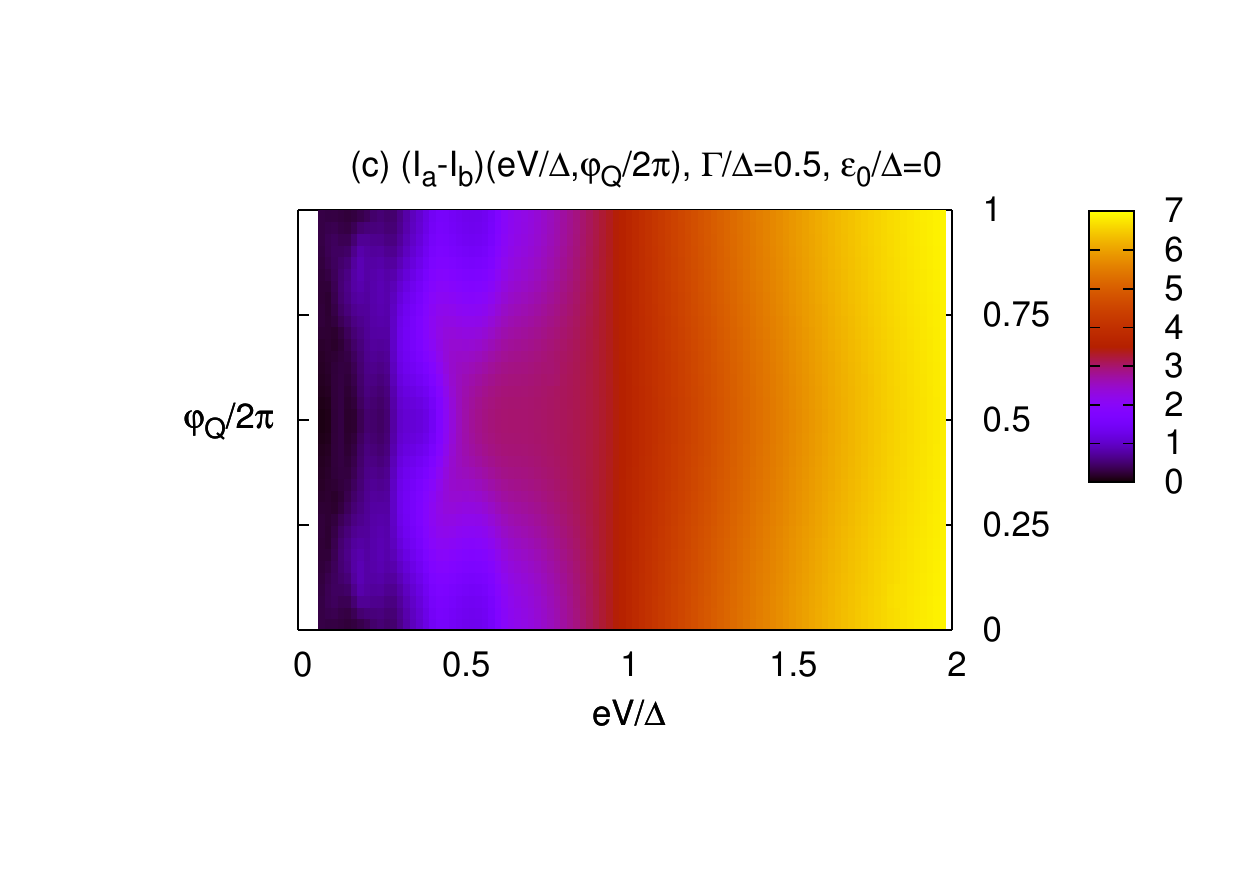}\includegraphics[width=.5\textwidth]{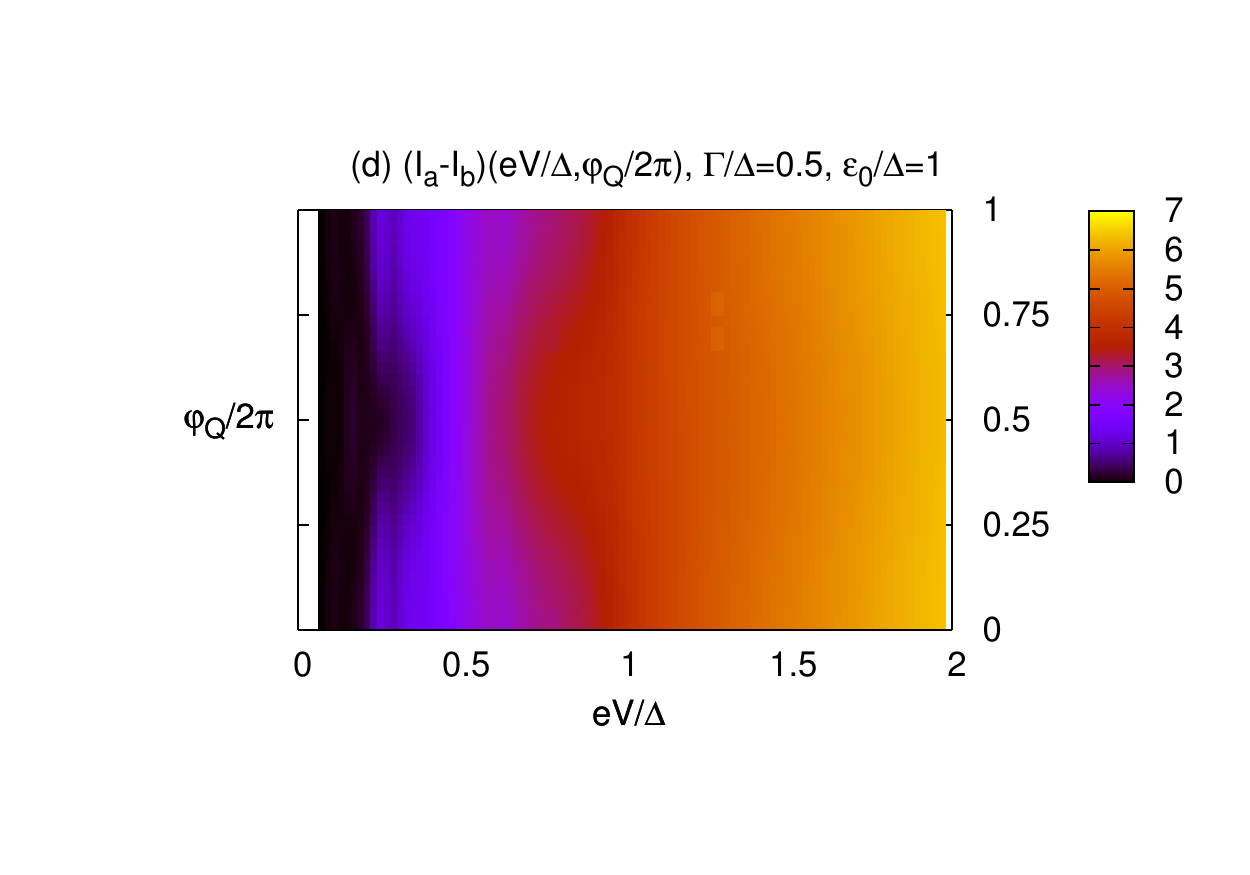}}
\vspace*{-1.5cm}
\centerline{\includegraphics[width=.5\textwidth]{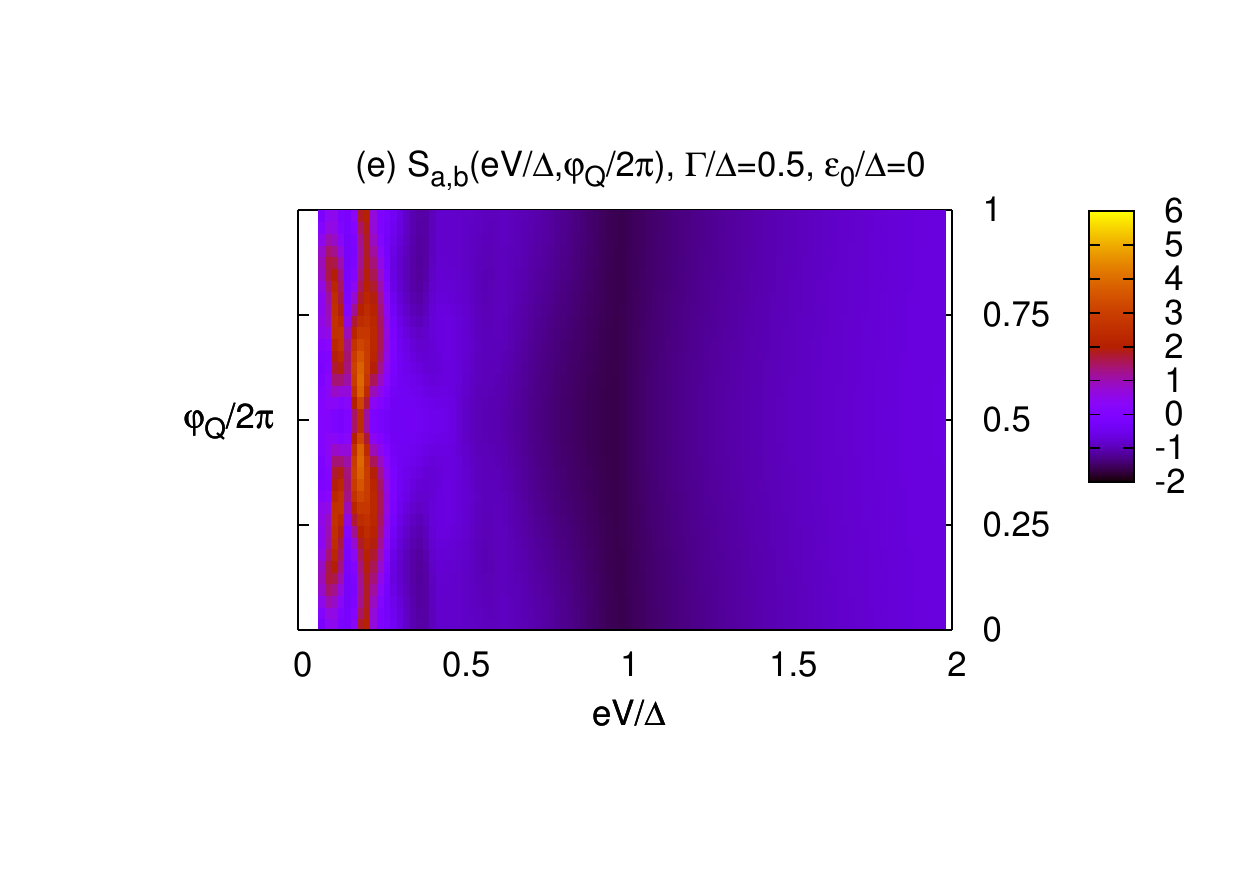}\includegraphics[width=.5\textwidth]{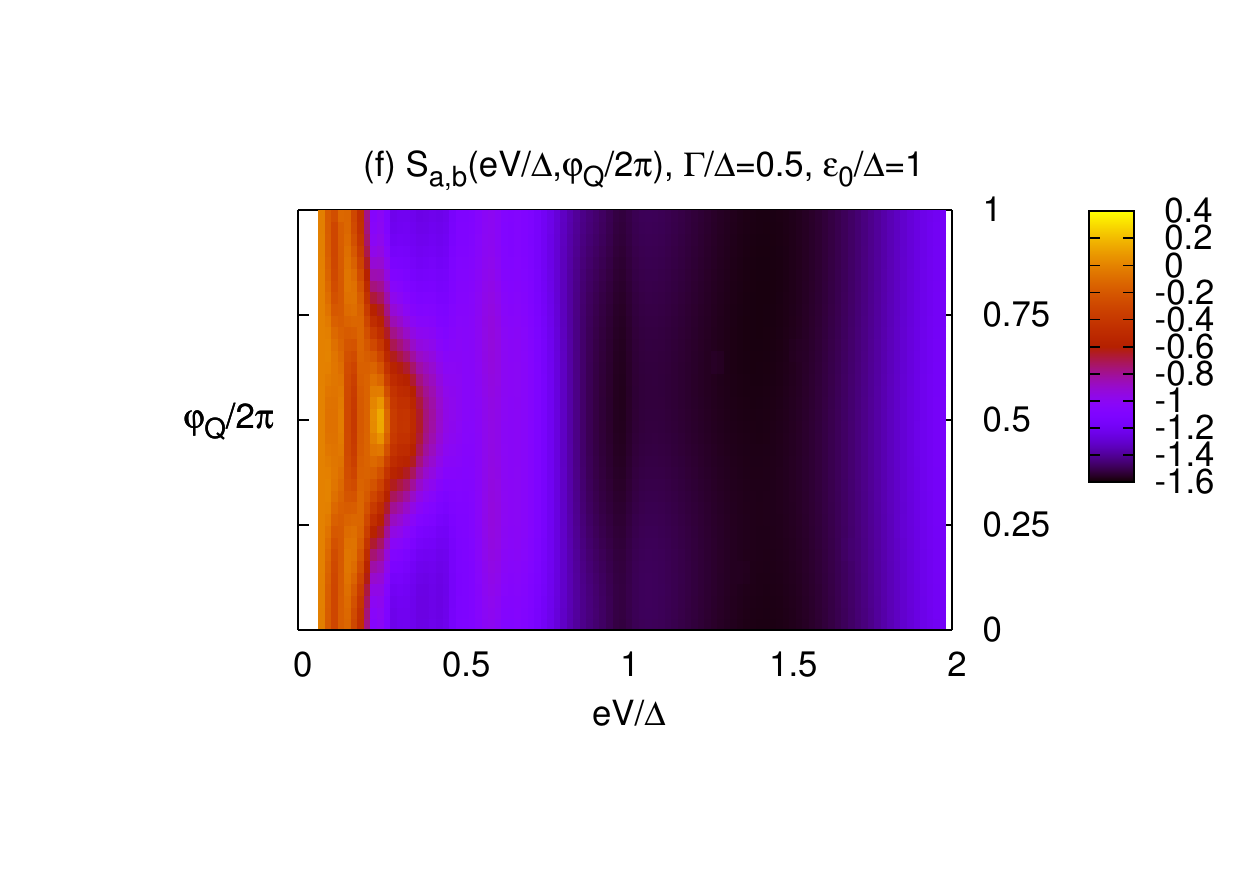}}
\caption{Color-map in the $(eV/\Delta ,\varphi_Q/2\pi)$ plane of the
  multipair current $I_c$ (panels a, b), of the current difference
  $I_a-I_b$ (panels c, d), and of the current cross-correlations
  $S_{a,b}$ (panels e, f).  The current is in units of $e\Delta/h$ and
  the current cross-correlations are un units of $e^2\Delta/h$. The
  contact transparencies are such that $\Gamma/\Delta=0.5$. Panels a,
  c and e correspond to $\epsilon_0/\Delta=0$, and panels b, d, f to,
  $\epsilon_0/\Delta=1$.
\label{fig:Ic_IamIb_Sab}
}
}
\end{figure*}

Now, numerical results are presented for the perturbative calculation
in transparency, in which the $28$ lowest-order terms in the quartet
contribution to the zero-frequency cross-correlations $S_{Q,ab}(0)$
are evaluated numerically at zero phase. The voltage dependence of
$S_{Q,ab}(0)$ is shown (in log scale) in Fig.~\ref{fig:pert-dep-V},
for different values of $\eta/\Delta$ over four orders of
magnitude. The small parameter $\eta\ll \Delta$ corresponds to a
line-width broadening introduced as the imaginary part to the energy,
and intended to regularize perturbation theory.  If $eV/\Delta>1$, the
$\varphi_Q=0$ cross-correlations are negative and large in absolute
value, due to the fact that, in this voltage range, extended
electron-like states below the gap of $S_a$ are coupled by the
quartets to extended hole-like states above the gap of $S_b$. As
$eV/\Delta$ is reduced below unity, much smaller values of $S_{Q,ab}$
are obtained, because $S_{Q,ab}$ is due to the residual density of
states inside the superconducting gap. Shoulders appear in the
voltage-dependence of the cross-correlations, due to the gap edge
singularities.  Extrapolating to $\eta/\Delta\rightarrow 0^+$ leads to
the conclusion that nonresonant-dot quartets do not contribute to the
current-current cross-correlations at subgap voltage.

\section{Quantum dot-superconductor three-terminal Josephson junction}
\label{sec:3T-Josephson-junction}

A few results are known for the current-current cross correlations in
a three-terminal all-superconducting structure with arbitrary
interface transparencies. Phase-insensitive positive
cross-correlations were discovered by Duhot, Lefloch and Houzet
\cite{Duhot} in the incoherent regime. The phase-sensitive thermal
noise and noise cross-correlations of a superconducting structure at
equilibrium was calculated by Freyn {\it et al.}\cite{Freyn} with the
Hamiltonian approach. Very recently, Riwar {\it et
  al.}\cite{Riwar-noise} provided a fully nonperturbative calculation
of the noise of a three-terminal Josephson junction biased at equal
voltages. In what follows, the junction is biased at opposite
voltages, therefore allowing for the emergence of a nonstandard
quartet mode, not present for equal voltages.

\begin{figure*}[htb]
\centerline{\includegraphics[width=.55\columnwidth]{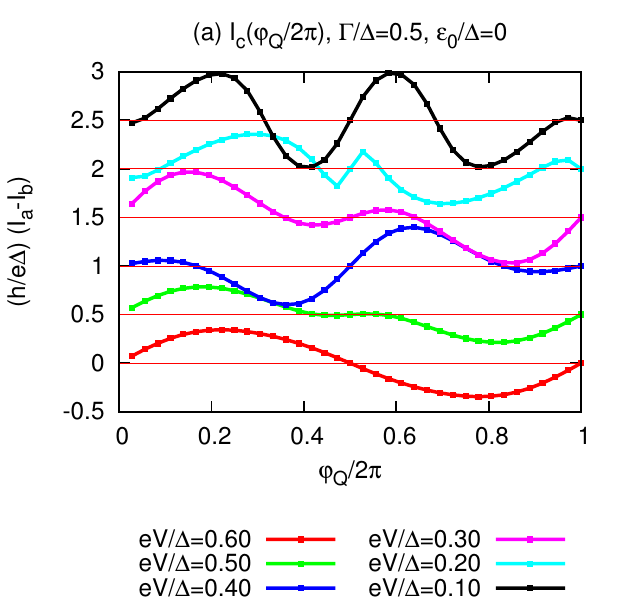}\includegraphics[width=.55\columnwidth]{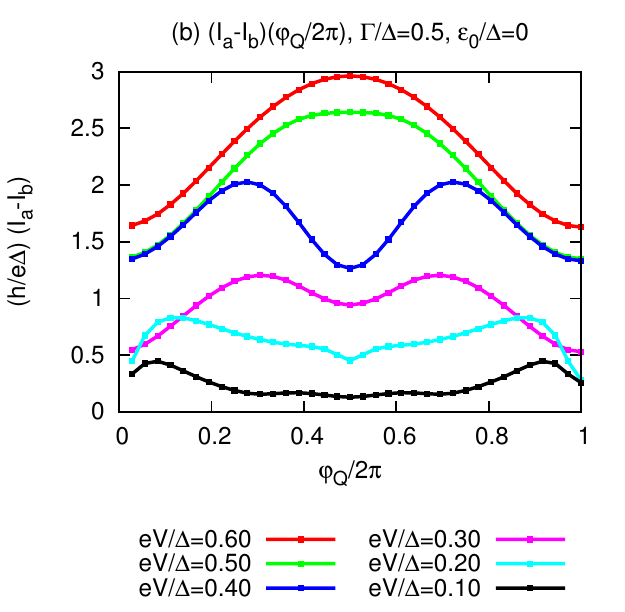}\includegraphics[width=.55\columnwidth]{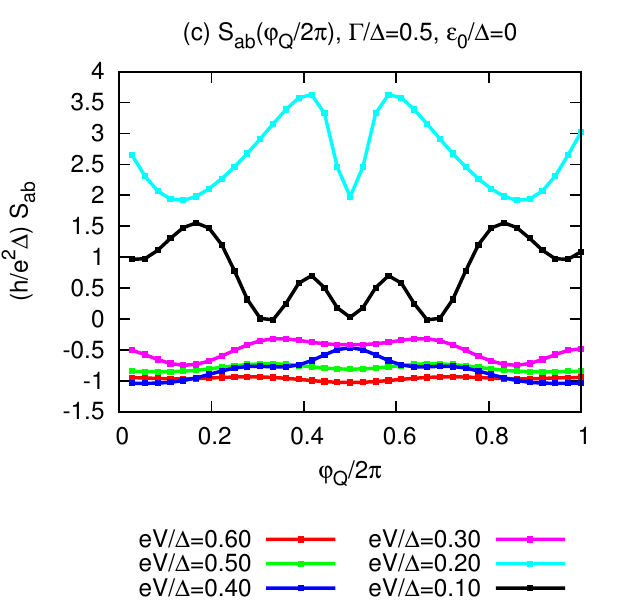}}
\centerline{\includegraphics[width=.55\columnwidth]{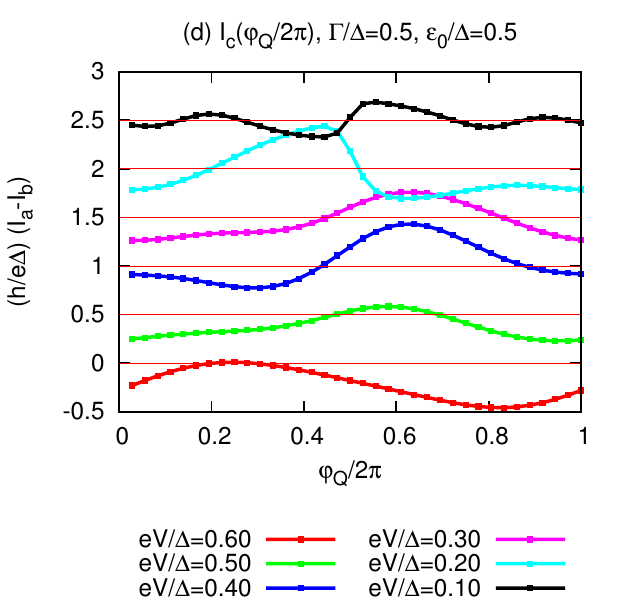}\includegraphics[width=.55\columnwidth]{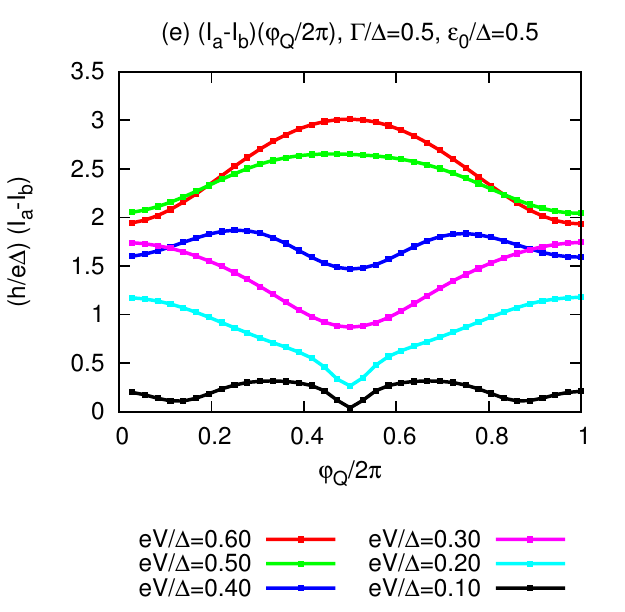}\includegraphics[width=.55\columnwidth]{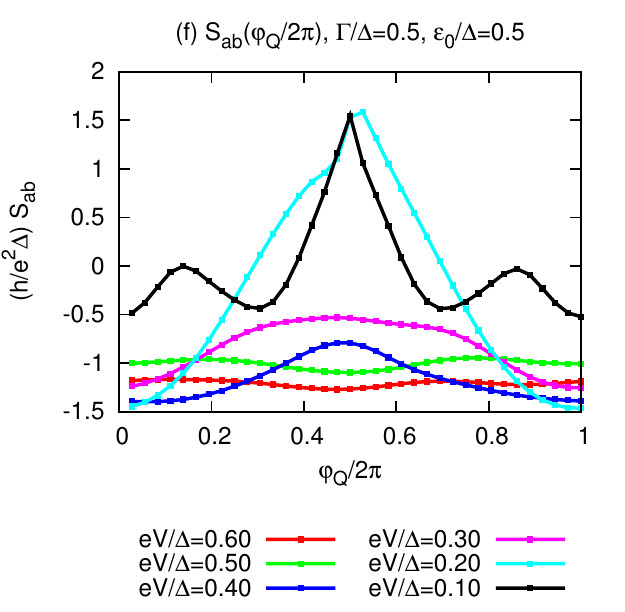}}
\centerline{\includegraphics[width=.55\columnwidth]{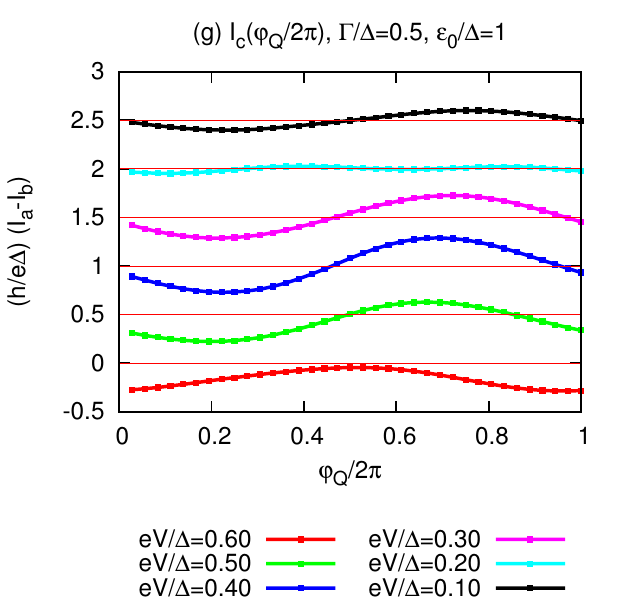}\includegraphics[width=.55\columnwidth]{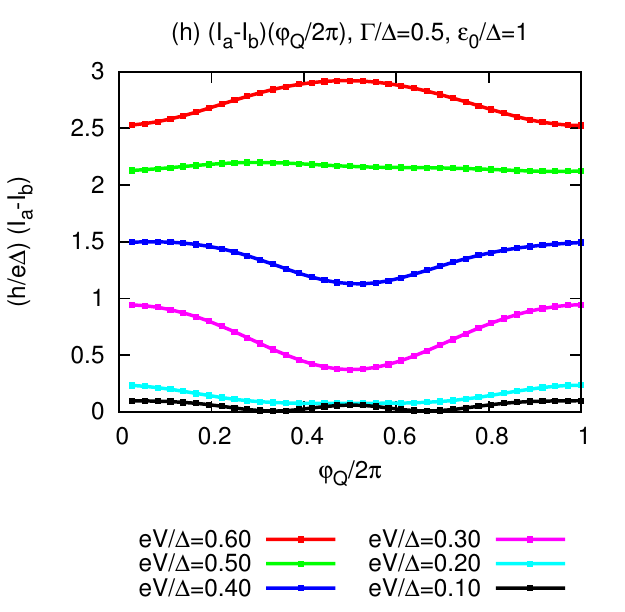}\includegraphics[width=.55\columnwidth]{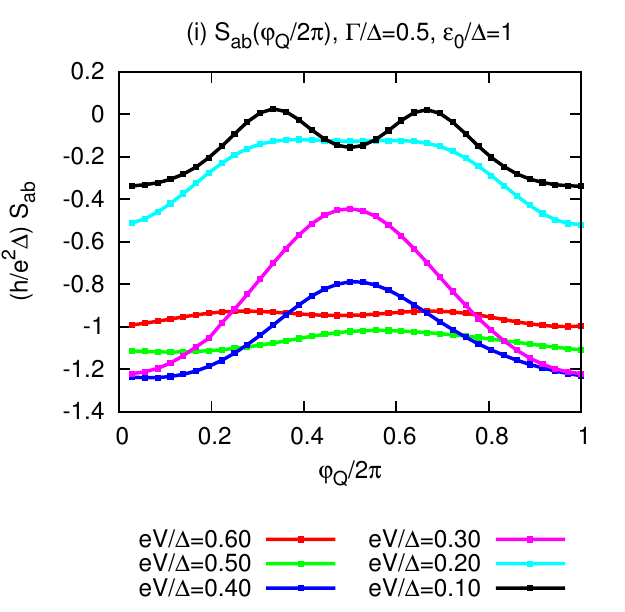}}
\centerline{\includegraphics[width=.55\columnwidth]{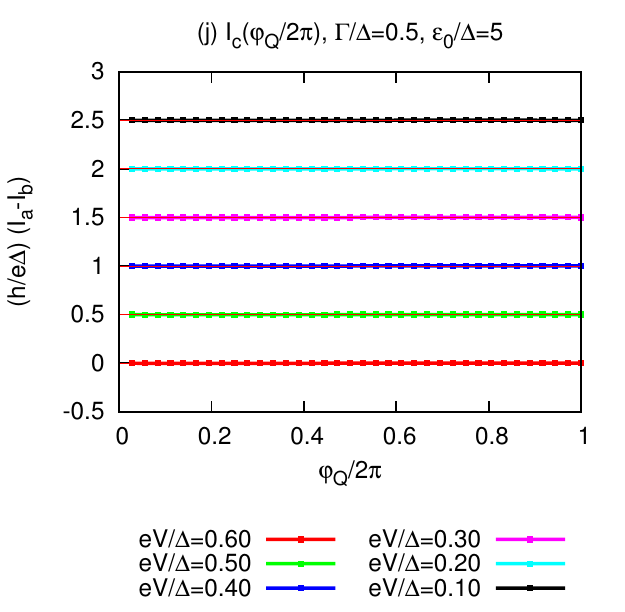}\includegraphics[width=.55\columnwidth]{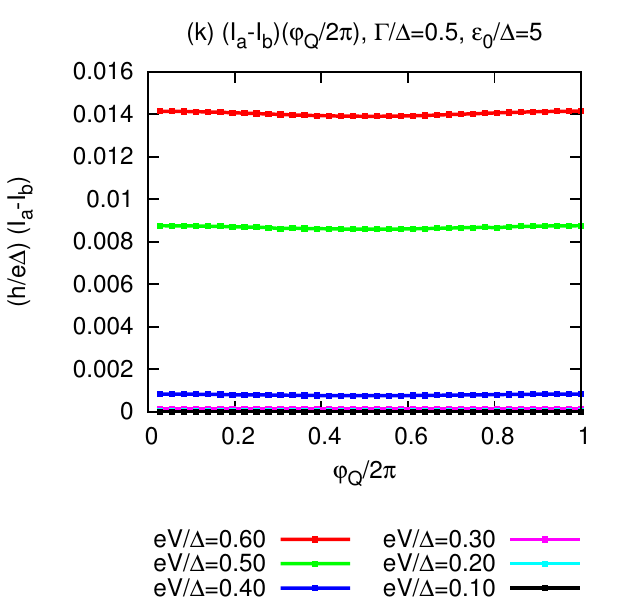}\includegraphics[width=.55\columnwidth]{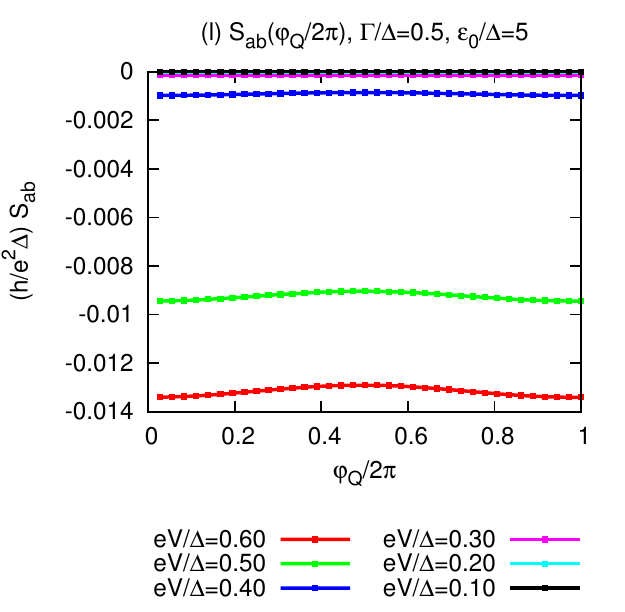}}
\caption{The figure shows the sensitivity on the phase $\varphi_Q$ of
  $I_c$ (panels a, d, g and j), $I_a-I_b$ (panel b, e, h and k) and
  $S_{a,b}$ (panels c, f, i and l), for the voltages indicated on the
  figure. For clarity, the data on
  panel a, d, g and j were shifted along $y$ axis according to the
  solid red lines. No shift is applied to the other panels.
\label{fig:I-phi}}
\end{figure*}

It is first recalled that a quantum dot is connected to three
superconducting leads $S_a$, $S_b$ and $S_c$ biased at opposite
voltages $V_a=-V_b\equiv V$, and $V_c=0$ respectively. The
normal-state transparency of the contacts is controlled by
$\Gamma=t^2/W$, where $t$ is the hopping amplitude between the dot and
the superconductors in their normal state, and $W$ is the hopping term
in the bulk of the superconductors (a fraction of the bandwidth). It
is supposed now that a single energy level is within the
superconducting gap window, and, in addition, this energy level
$\epsilon_0$ (controllable by a gate voltage) is varied
systematically, thus allowing to cross-over from nonresonant-dot quartets
(for $\epsilon_0/\Delta\gg 1$) to resonant-dot quartets for
$\epsilon_0/\Delta\alt 1$, with different behavior of the noise in
both regimes.

It was established by Jonckheere {\it et al.}\cite{Jonckheere} that
the current has two components: with particle-hole symmetry, the
current $I_c$ (due to multipairs generalizing quartets) is even in
voltage and odd {in the phase $\varphi_Q$}, and the current
difference $I_a-I_b$ (due to ph-MARs) is odd in voltage and even
{in the phase $\varphi_Q$}. {Fig.~\ref{fig:Ic_IamIb_Sab}
  shows how $I_c$, the current difference $I_a-I_b$ and the
  cross-correlations $S_{a,b}$ vary in the parameter plane
  $(eV/\Delta,{\varphi_Q}/2\pi)$, for the experimentally relevant
  intermediate $\Gamma/\Delta=0.5$.} {The current and noise {
    exhibit} a dependence} on the three-body phase variable
$\varphi_Q=\varphi_a+\varphi_b-2\varphi_c$.  {Panels a, c, e and
  b, d, f of Fig.~\ref{fig:Ic_IamIb_Sab} correspond respectively to
  $\epsilon_0/\Delta=0$ and {$\epsilon_0/\Delta=1$}, thus in the
  resonant dot regime.} {The values of the current and noise
  cross-correlations are large in the resonant dot regime
  $\epsilon_0/\Delta\alt 1$, which contrasts with the nonresonant dot
  regime $\epsilon_0/\Delta\gg 1$ (see the preceding
  Sec.~\ref{sec:perturbative-quartets}).} The current $I_c$, the
current difference $I_a-I_b$ and the cross-correlations $S_{a,b}$ have
a strong dependence on the {quartet} phase $\varphi_Q$ in the
nonresonant dot regime $\epsilon_0/\Delta\alt 1$. {A weak dependence
  on $\varphi_Q$ of those quantities was obtained numerically for
  $\epsilon_0/\Delta=5$ (not shown in Fig.~\ref{fig:Ic_IamIb_Sab}), in
  a qualitative agreement with Sec.~\ref{sec:perturbative-quartets}.}

The current-current cross-correlations $S_{a,b}$ are shown by a
color-plot in {Fig.~\ref{fig:Ic_IamIb_Sab}e and
  Fig.~\ref{fig:Ic_IamIb_Sab}f} in the plane {of the variables}
$(eV/\Delta,\varphi_Q/2\pi)$, for the same values
$\epsilon_0/\Delta=0$ (panel e) and {$\epsilon_0/\Delta=1$} (panel
f). Positive and phase-sensitive current-current cross-correlations
resonances emerge below $eV/\Delta\alt 0.4$. An experiment {
  measuring} cross-correlations in the $(V_a,V_b)$ plane should {
  thus} detect an additional contribution to the cross-correlations if
the three-body phase variable $\varphi_Q$ becomes a relevant quantity
at the quartet resonance $V_a+V_b=0$ (with $V_c=0$).

\begin{figure*}[htb]
\centerline{\includegraphics[width=.32\textwidth]{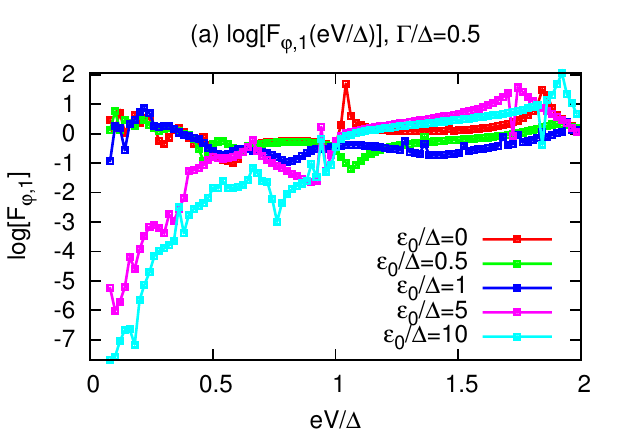}\includegraphics[width=.32\textwidth]{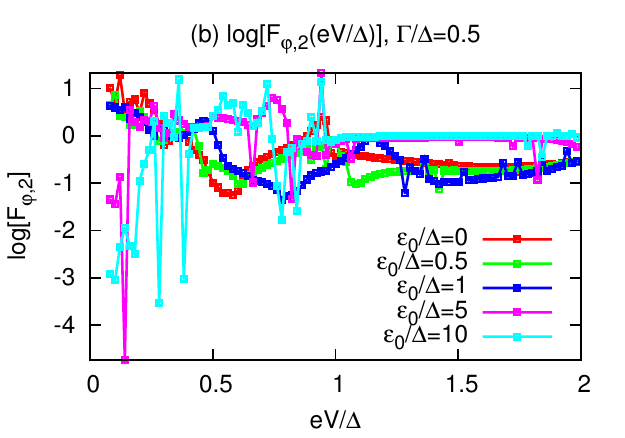}\includegraphics[width=.32\textwidth]{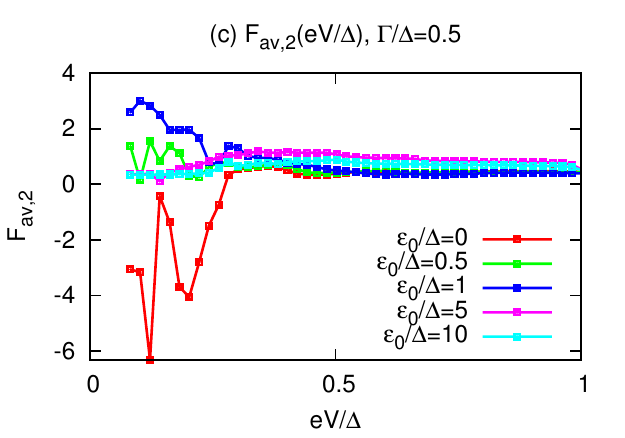}}
\caption{The figure shows the sensitivity on normalized voltage
  $eV/\Delta$ of the logarithm of the phase Fano factors
  $F_{\varphi,1}=\delta S_{a,b}/\delta (I_c)$ (panel a) and
  $F_{\varphi,2}=\delta S_{a,b}/\delta (I_a-I_b)$ (panel b), where the
  symbol $\delta X$ has the meaning of $\delta
  X=\mbox{Max}_{\varphi_Q} X(\varphi_Q)-\mbox{Min}_{\varphi_Q}
  X(\varphi_Q)$. Panel c shows the $eV/\Delta$-dependence of the
  average Fano factor $F_{av,2}= S_{a,b,av}/I_{c,av}$,
  where the subscript ``av'' denotes averaging over the phase
  $\varphi_c$.
\label{fig:F-phi}}
\end{figure*}

\begin{figure*}[htb]
\centerline{\includegraphics[width=\columnwidth]{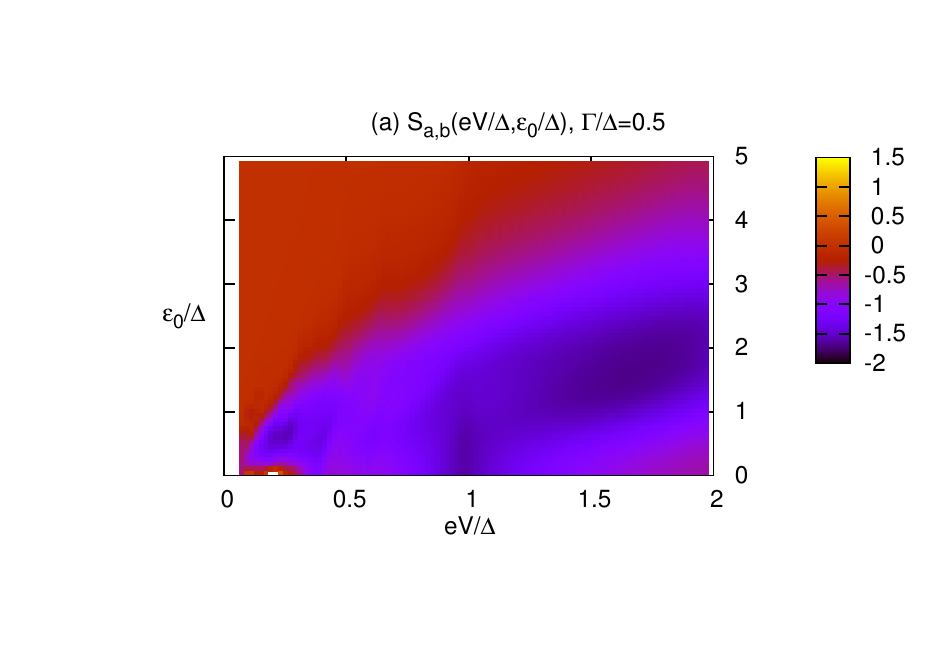}\includegraphics[width=\columnwidth]{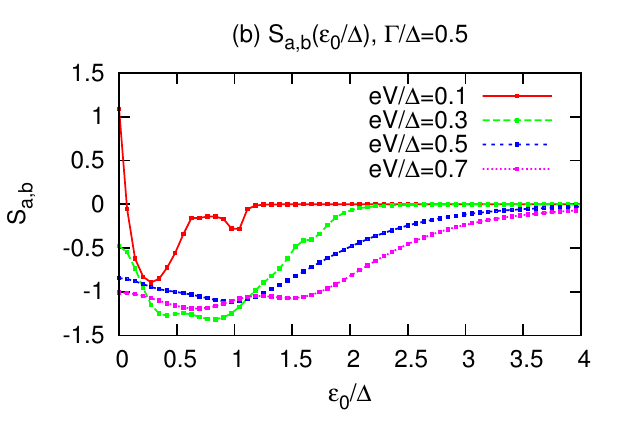}}
\caption{Current-current cross-correlations
  $S_{a,b}(eV/\Delta,\epsilon_0/\Delta)$ for $\Gamma/\Delta=0.5$
  (panel a). The cross-correlations become negligible in the
  nonresonant dot regime $\epsilon_0/\Delta\agt \epsilon_0^*/\Delta$. The
  value of $\epsilon_0^*/\Delta$ decreases to zero as $eV/\Delta$ is
  reduced. Panel b shows the current-current cross-correlations
  $S_{a,b}(\epsilon_0/\Delta)$ for $\Gamma/\Delta=0.5$, and for the
  values $eV/\Delta=0.1, 0.3, 0.5, 0.7$.\label{fig:Sab-Veps}}
\end{figure*}

The color-plots in {Fig.~\ref{fig:Ic_IamIb_Sab} for $I_c$,
  $I_a-I_b$ and $S_{a,b}$} are {complemented by conventional
  one-parameter plots (see Fig.~\ref{fig:I-phi}) which better
  illustrate the phase sensitivity of the currents and current-current
  cross-correlations. Let us first consider the multipair current
  (Fig.~\ref{fig:I-phi} a, d, g, j), which is, as expected, odd in the
  phase $\varphi_Q$. A strongly anharmonic behavior is clearly
  obtained for $\epsilon_0/\Delta\alt 1$ and $eV/\Delta \ll 1$, with a
  quasi-period doubling as $eV/\Delta$ is reduced from $eV/\Delta=0.6$
  to $eV/\Delta=0.1$ if $\epsilon_0/\Delta=0$ (Fig.~\ref{fig:I-phi}
  a), pointing towards emerging octets at low bias.  {
    Quasi-harmonic and ``$0$''-junction behaviors are recovered} for
  {vanishingly small} $\epsilon_0/\Delta=0$ and larger
  $eV/\Delta$. In contrast, for larger $\epsilon_0/\Delta$, an
  harmonic behavior is obtained with a ``$\pi$''-junction
  character. Second, the quasiparticle current $I_a-I_b$ is, as
  expected, even in phase, and, contrarily to $I_c$, it has a nonzero
  phase-averaged value (Fig.~\ref{fig:I-phi} b, e, h, k).  The latter
  represents the ``usual'' phase-insensitive MARs, which increases
  with $eV/\Delta$.  On the other hand, the phase modulation
  represents the phase-MARs and it also displays anharmonic behavior at
  small voltage. Third, {the panels c, f, i and l of
    Fig.~\ref{fig:I-phi}} represent the cross-correlations
  $S_{a,b}(\varphi_Q)$. As a new result, one finds that, like the
  quasiparticle current, it is even in phase, and it has a nonzero
  phase average}.  An especially complex harmonic content is obtained
on panel c.  A general trend is that negative current-current
cross-correlations are obtained for $\epsilon_0/\Delta=5$, which
become negligibly small as the voltage is reduced below $eV/\Delta\alt
0.3$ (see Fig.~\ref{fig:I-phi}l). This behavior is consistent with the
absence of current-current cross-correlations for the nonresonant-dot
quartets at low bias voltage (see
Sec.~\ref{sec:perturbative-quartets}). Positive current-current
cross-correlations emerge gradually as $\epsilon_0/\Delta$ is reduced,
first for the lowest bias voltage $eV/\Delta=0.1$ in a specific window
of the phase variable $\varphi_Q$ if $\epsilon_0/\Delta=1$ (see
Fig.~\ref{fig:I-phi}i). Positive current-current cross-correlations
are obtained for the lowest value $\epsilon_0/\Delta=0$ (see
Fig.~\ref{fig:I-phi}c), at low normalized bias voltage
$eV/\Delta=0.1\div 0.2$ and in the full range of $\varphi_Q$.

A closer look at panels a-l of Fig.~\ref{fig:I-phi} reveals that the
current-current cross-correlations correlate weakly with the multipair
current $I_c$, but the correlation is better with the current
difference $I_a-I_b$ (corresponding to the physical process of
ph-MARs). One notices that ``kinks'' emerge in $S_{a,b}$ at
$\varphi_c=\pi/2$ for $\epsilon_0/\Delta=0.5$ and $eV/\Delta=0.1$,
$0.2$ (see Fig.~\ref{fig:I-phi}f). Those kinks in the
cross-correlations are to be put in correspondence with similar
features in $I_a-I_b$ (ph-MARs, see Fig.~\ref{fig:I-phi}e), not
present in $I_c$ (multipair current, see Fig.~\ref{fig:I-phi}d). The
same analogy between $S_{a,b}$ and $I_a-I_b$ is also visible for
$\epsilon_0/\Delta=0$ (see Figs.~\ref{fig:I-phi}a, b and c). 

It is relevant both experimentally and theoretically to compare the
values of the cross-correlations to the values of the currents. {
  It was found previously (see Fig.~\ref{fig:Ic_IamIb_Sab}) that the
  cross-correlations become very small in the nonresonant dot regime
  $\epsilon_0/\Delta\gg 1$ and in the limit of low bias voltage
  $eV/\Delta\ll 1$. However, the phase-sensitive current is also
  reduced if $\epsilon_0/\Delta\gg 1$, and the question arises of
  comparing the noise to the current in the nonresonant dot regime at low
  bias voltage.} The quantity $\delta S_{a,b}$ is defined as the
difference between the maximum and the minimum (over the phase
$\varphi_Q$) of $S_{a,b}(\varphi_Q)$, and a similar definition holds
for $\delta I_c$ and $\delta [I_a-I_b]$. A first Fano factor is
defined as $F_{\varphi,1}=\delta S_{a,b}/\delta I_c$, which is the
value of the amplitude of the oscillations of the cross-correlations
normalized to the amplitude of the oscillations of the multipair
current $I_c$. [The symbol $\delta X=\mbox{Max}_{\varphi_Q}
  X(\varphi_Q)-\mbox{Min}_{\varphi_Q} X(\varphi_Q)$ has the meaning of
  an amplitude phase variations.] The second Fano factor is defined as
the amplitude of the oscillations of the cross-correlations normalized
to that of the phase-MAR processes: $F_{\varphi,2}=\delta
S_{a,b}/\delta [I_a-I_b]$.  The voltage dependence of
$\log(F_{\varphi,1})$ and $\log(F_{\varphi,2})$ are shown in
Figs.~\ref{fig:F-phi}a and b respectively. The different curves on
each of those panels correspond to the values $\epsilon_0/\Delta=0$,
$0.5$, $1$, $5$ and $10$. The spikes on panel b correspond to values
of the voltage for which the integral over energy $\omega$ of the
current is very small, therefore deteriorating the accuracy in the
Fano factor $F_{\varphi,2}$. Indeed, it turns out that, for specific
voltages, the amplitudes of oscillations in $I_a-I_b$ can become very
small, because the difference $I_a-I_b$ goes to zero in the
zero-voltage limit. The data-points shown on panels a and b of
Fig.~\ref{fig:F-phi} correspond to unsmoothed raw data that are
however sufficient for the purpose of discussing now the general
trends. If $\epsilon_0/\Delta=5$, $10$, the Fano factors
$F_{\varphi,1}$ and $F_{\varphi,2}$ decrease drastically towards zero
as $eV/\Delta$ is reduced. If $\epsilon_0/\Delta=0$, $0.5$, $1$ the
Fano factors $F_{\varphi,1}$ and $F_{\varphi,2}$ take much higher
values of order $0.1\div 1$. {In addition, the Fano factor for
  the noise and current averaged over the phases
  $F_{av,2}=S_{a,b,av}/[I_{a,av}-I_{b,av}]$ is shown on panel c of
  Fig.~\ref{fig:F-phi}, which also demonstrates a strong reduction of
  $|F_{av,2}|$ at low bias in the nonresonant dot regime}. [The symbol
  $X_{av}=\int X(\varphi_Q) d\varphi_Q/2\pi$ has the meaning of an
  average over $\varphi_Q$.] In addition, a nontrivial change of sign
is obtained in $F_{av,2}$, which reflect the overall sign of $S_{a,b}$
(see Fig.~\ref{fig:Ic_IamIb_Sab}, Fig.~\ref{fig:I-phi} and the
forthcoming Fig.~\ref{fig:Sab-Veps}b).

The results presented in Fig.~\ref{fig:F-phi} demonstrate that, at low
bias voltage, the cross-correlations $S_{a,b}$ tend to zero faster
than the currents in the strongly nonresonant dot regime
$\epsilon_0/\Delta\gg 1$. The cross-correlations $S_{a,b}$ (in units
of the currents) thus become very small in the nonresonant dot regime, but not
in the resonant dot regime, suggesting that a gate voltage can be
used to monitor the value of cross-correlations at the quartet
resonance $V_a+V_b=0$.

The crossover between the resonant and nonresonant dot regimes is better
visualized in Fig.~\ref{fig:Sab-Veps}a and b. Fig.~\ref{fig:Sab-Veps}a
shows in color-scale the value of the cross-correlations in the plane
$(eV/\Delta,\epsilon_0/\Delta)$, for $\Gamma/\Delta=0.5$ and
$\varphi_Q=0$. The red area in the top-left corner
of Fig.~\ref{fig:Sab-Veps}a corresponds to the nonresonant dot regime in
which the cross-correlations are very small. The blue area corresponds
to large negative cross-correlations. The positive cross-correlations
are restricted to the bottom-left corner, as seen from panel b showing
the current-current cross-correlations as a function of
$\epsilon_0/\Delta$ for different values of $eV/\Delta$. At fixed
$eV/\Delta$, there is thus a cross-over value $\epsilon_0^*/\Delta$ of
the parameter $\epsilon_0/\Delta$ above which the cross-correlations
are weak. The value of $\epsilon_0^*/\Delta$ is strongly reduced as
the normalized voltage $eV/\Delta$ is reduced, which appears to be
compatible with the absence of current-current cross-correlations in
the adiabatic limit (see Sec.~\ref{sec:adiabatic}).

\section{Conclusions}
\label{sec:conclusions}
To conclude, it is a relevant question to ask whether splitting a
supercurrent {by quartets at resonant voltages} 
produces positive cross-correlations at zero
temperature. Splitting a supercurrent at equilibrium or in the
adiabatic limit does not produce noise, and our numerical calculations
are consistent with this limit of low bias voltage. It was shown by a
semi-analytical perturbative calculation in interface transparency
that the quartets are noiseless also in the nonresonant dot regime in the
limit of small interface transparencies, for arbitrary voltage below
the gap. Those perturbative calculations in interface transparency
took the full Keldysh structure into account. However, phase-sensitive
positive current-current cross-correlations are obtained numerically
in the resonant dot case. A quantum dot was connected to three
superconductors with an intermediate coupling $\Gamma/\Delta=0.5$. The
resonant dot regime was obtained if the quantum dot energy level
$\epsilon_0$ is such that $\epsilon_0/\Delta\ll 1$ and the nonresonant dot
regime corresponds to $\epsilon_0/\Delta\gg 1$. These phase-sensitive
current cross-correlations correlate in a qualitative manner with the
signal of phase-sensitive MARs, which suggests {a strong contribution from the latter}. Those
phase-sensitive MARs correspond to the transmission of a quasiparticle
assisted by quartets or by multipairs. In this respect, a nonzero value
for the phase-sensitive component of current cross-correlations in
 noise experiments would imply that quartets or multipairs
are present together with quasiparticles. One can conclude that a
cross-correlation experiment should detect a gate-tunable anomaly at
the quartet resonance $V_a+V_b=0$. A strong phase-sensitivity of the
cross-correlations is predicted in the resonant dot regime
$\epsilon_0/\Delta\alt 1$, and an absence of noise cross-correlations
is obtained in the nonresonant dot regime $\epsilon_0/\Delta\gg 1$. In an
experiment, the width in voltage $(V_a,V_b)$ parameter plane of the
anomaly obtained for $\epsilon_0/\Delta\alt 1$ in the
cross-correlations is expected to correlate to the Josephson anomaly
in the average current, because both anomalies
originate from the appearance of the three-body phase variable
$\varphi_Q=\varphi_a+\varphi_b-2\varphi_c$. It is noted finally that
phase-sensitive noise was already calculated and measured in Andreev
interferometers \cite{Nazarov,Heikkila}. It is proposed here to go one
step further and measure an anomaly in the noise or in the
cross-correlations of a three-terminal Josephson junction.

\section*{Acknowledgements}
The authors acknowledge financial support from the French ``Agence
National de la Recherche'' under contract ``Nanoquartets''
12-BS-10-007-04. The authors thank the CRIHAN computing center for the
use of its facilities. R.M. used the computer facilities of Institut
N\'eel to develop and test the codes for the current and for the
noise.  The authors acknowledge fruitful discussions with
T. Jonckheere, T. Martin, J. Rech, and with H. Courtois, M. Houzet and
J. Meyer. The authors wish to express special thanks to F. Lefloch for
his encouragements and {suggestions}. R.M. and D.F. wish to acknowledge an useful
correspondence with Y. Cohen, M. Heiblum and Y. Ronen.

\appendix

\section*{Recursive Green's functions in energy for three-terminal
structures}

This Appendix generalizes to three terminals the algorithm proposed by
Cuevas, Mart\'{\i}n Rodero and Levy Yeyati \cite{Cuevas,Cuevas-noise}
in which the current of MARs was evaluated in a two-terminals
structure. All numerical calculations for the three-terminal junction
were realized on the basis of of this method.

The Green's functions $\hat{G}_{n,m}(\omega)$ depend on one energy
$\omega$ and two integers $n$ and $m$ (the harmonics of half the
Josephson frequency). The Dyson equation takes the form
\begin{eqnarray}
\label{eq:G-tb}
\hat{G}_{n,m}&=&\hat{K}_{n,n} \hat{G}_{n,m} +\hat{K}_{m,m}^{(0)}\delta_{n,m}
+\hat{K}_{n,n+2} \hat{G}_{n+2,m} \\&+& \hat{K}_{n,n-2} \hat{G}_{n-2,m}
\nonumber
,
\end{eqnarray}
where the dependence on $\omega$ is made implicit. The matrices $K$
have three component, one for each of the terminals:
\begin{widetext}
\begin{eqnarray}
\hat{K}^{n,n}_{(a)}&=&\left(\begin{array}{cc} g_{x,x}^{1,1/n,n}\Sigma_{x,a}^{1,1/n,n+1}
g_{a,a}^{1,1/n+1,n+1}\Sigma_{a,x}^{1,1/n+1,n} & 0 \\
0 & g_{x,x}^{2,2/n,n}\Sigma_{x,a}^{2,2/n,n-1} g_{a,a}^{2,2/n-1,n-1} \Sigma_{a,x}^{2,2/n-1,n}
\end{array}\right)\\
\hat{K}^{n,n+2}_{(a)}&=&\left(\begin{array}{cc} 0 & g_{x,x}^{1,1/n,n}\Sigma_{x,a}^{1,1/n,n+1}
g_{a,a}^{1,2/n+1,n+1}\Sigma_{a,x}^{2,2/n+1,n+2} \\
0 & 0
\end{array}\right)\\
\hat{K}^{n,n-2}_{(a)}&=&\left(\begin{array}{cc} 0 & 0\\
g_{x,x}^{2,2/n,n}\Sigma_{x,a}^{2,2/n,n-1}
g_{a,a}^{2,1/n-1,n-1}\Sigma_{a,x}^{2,2/n-1,n-2} & 0
\end{array}\right).
\end{eqnarray}
Similar expressions are obtained for $K_{(b)}$
\begin{eqnarray}
\hat{K}^{n,n}_{(b)}&=&\left(\begin{array}{cc} g_{x,x}^{1,1/n,n}\Sigma_{x,b}^{1,1/n,n-1}
g_{b,b}^{1,1/n-1,n-1}\Sigma_{b,x}^{1,1/n-1,n} & 0 \\
0 & g_{x,x}^{2,2/n,n}\Sigma_{x,b}^{2,2/n,n+1} g_{b,b}^{2,2/n+1,n+1} \Sigma_{b,x}^{2,2/n+1,n}
\end{array}\right)\\
\hat{K}^{n,n-2}_{(b)}&=&\left(\begin{array}{cc} 0 & g_{x,x}^{1,1/n,n}\Sigma_{x,b}^{1,1/n,n-1}
g_{b,b}^{1,2/n-1,n-1}\Sigma_{b,x}^{2,2/n-1,n-2} \\
0 & 0
\end{array}\right)\\
\hat{K}^{n,n+2}_{(b)}&=&\left(\begin{array}{cc} 0 & 0\\
g_{x,x}^{2,2/n,n}\Sigma_{x,b}^{2,2/n,n+1}
g_{b,b}^{2,1/n+1,n+1}\Sigma_{b,x}^{2,2/n+1,n+2} & 0
\end{array}\right).
\end{eqnarray}
and for $K_{(c)}$:
\begin{eqnarray}
\hat{K}^{n,n}_{(c)}&=&\left(\begin{array}{cc} g_{x,x}^{1,1/n,n}\Sigma_{x,c}^{1,1/n,n}
g_{c,c}^{1,1/n,n}\Sigma_{c,x}^{1,1/n,n} & 
g_{x,x}^{1,1/n,n}\Sigma_{x,c}^{1,1/n,n} g_{c,c}^{1,2/n,n} \Sigma_{c,x}^{2,2/n,n}\\
g_{x,x}^{2,2/n,n}\Sigma_{x,c}^{2,2/n,n}
g_{c,c}^{2,1/n,n}\Sigma_{c,x}^{1,1/n,n} & 
g_{x,x}^{2,2/n,n}\Sigma_{x,c}^{2,2/n,n} g_{c,c}^{2,2/n,n} \Sigma_{c,x}^{2,2/n,n}
\end{array}\right)\\
\hat{K}_{(c)}^{n,n+2}&=&K_{(c)}^{n,n-2}=0
\end{eqnarray}
The matrix $\hat{K}^{(0)}$ is as follows:
\begin{eqnarray}
\hat{K}^ {(0)m,m}=\left(\begin{array}{cc} g_{x,x}^{1,1/m,m} & 0\\0 & g_{x,x}^{2,2/m,m}
\end{array}\right)
.
\end{eqnarray}
Next, Eq.~(\ref{eq:G-tb}) is solved by recursion:
$\hat{G}_{n-2,m}=\hat{z}^-_{n-2,m}\hat{G}_{n,m}$ leads to
\begin{equation}
\hat{z}^-_{n,n+2}=\left(\hat{I}-\hat{K}_{n,n}-\hat{K}_{n,n-2}\hat{z}^-_{n-2,n}\right)^{-1}\hat{K}_{n,n+2}
\end{equation}
for $n<m$. On the other hand, $\hat{G}_{n,m}=z^+_{n,n-2} \hat{G}_{n-2,m}$ leads to
\begin{equation}
\hat{z}^+_{n,n-2}=\left(\hat{I}-\hat{K}_{n,n}-\hat{K}_{n,n+2}\hat{z}^-_{n+2,n}\right)^{-1}\hat{K}_{n,n-2}
\end{equation}
if $n>m$. For $n=m$, we find
\begin{equation}
\hat{G}_{m,m}=\left(\hat{I}-\hat{K}_{m,m}-\hat{K}_{m,m+2} \hat{z}^+_{m+2,m}
-\hat{K}_{m,m-2}\hat{z}^-_{m-2,m}\right)^{-1}\hat{K}^{(0)}_{m,m}
.
\end{equation}
\end{widetext}

\end{document}